\documentclass[a4paper,12pt,fleqn]{article}
\usepackage{graphicx}
\usepackage{amssymb}
\usepackage{amsfonts}
\usepackage{color}
\usepackage{amsmath}
\usepackage{tikz}

\usepackage{graphicx,array}
\usepackage{amssymb,amsmath}

\def\const{\hbox{const}}
\def\sgn{\hbox{sgn}}
\newtheorem{theorem}{Theorem}

\title{Stabilization of the wheeled inverted pendulum on a soft surface}
\author{O.M.~Kiselev
}

\begin{document}

\maketitle

\begin{abstract}
We study dynamics of an wheeled inverted  pendulum under a proportional-integral-derivative controller on  horizontal, inclined and soft surfaces. An oscillatory area and conditions of the stability for the control are shown on the phase portraits of the dynamical systems. Particularly, we study a differential inclusion for  moving on the soft surface, and we find semi-stable stationary solutions in our mathematical model. Due to rounding errors of the numerical modelling or external perturbations of robotics equipment the semistability looks as a  limit cycle in simulations.
\end{abstract}

\section{Introduction}

The wheeled inverted pendulum (WIP) is the important model to study the robotic dynamics and control.  Contemporary works concerning this subject consider a wide diversity of the problems from the accurate mathematical models and to electrical scheme for the controls \cite{PathakFranchAgrawal2005}, \cite{SamiMichalskaAngeles2007}. But really, a backbone of  all such approaches is the simple mathematical model which contains most of  the constraints for stabilization and driving of the inverted wheeled pendulum as a robotic equipment \cite{Formalskii2016Eng}, \cite{MartynenkoFormalskii2013Eng}. 

The historic review of studying for the problem of controlling of the inverted pendulum in a lot of cases one can see in\cite{AndrievskiiFradkov1999Eng}, \cite{BlockAstromSpong2007}. The reviews of controlling of the WIP can be find in \cite{ChanStolHalkyard2013},  \cite{SilvaFrankSup2016}.

The discussion concerning stabilizing and controlling the WIP were examined in 
\cite{GrasserDArrigoColombiRufer2002},  \cite{PathakFranchAgrawal2005}, \cite{FDGVKP2017} and the same for inclined surface in \cite{SamiMichalskaAngeles2007}. The questions of moving with the vertical obstacles were discussed in  \cite{TeeyapanWangKunzStilman2010}.The movement on the uneven surfaces was studied in  \cite{KausarStolPatel2010}, \cite{KausarStolPatel2012}.

The proportional-integral-derivative controller is wide used in thechnics, see \cite{AstromHagglund1994}.  The general approaches for the linear model in the form of  the second order differential equations with firrst integrals one can see in for example in  \cite{Kiselev2019bookEng}.  The PID controllers is often used for two whilled robots, see  \cite{HatakeyamaShimada2008},\cite{NasirRajaIsmailAhmad2010}, and review \cite{ChanStolHalkyard2013}. 
The separated control and used the Lyapunov function for the WIP was considered for example in  \cite{MaddahiShamekhiGhaffari2015}.

In this work we derive the equations for the dynamics of the WIP on the uneven curve. We consider the different cases of such equations for the horizontal, for the inclined line and finally for the soft line.  We obtain the areas of the phase spaces where the WIP may be stabilized by the proportional-derivative controller. We show the change of the area of the stability for the WIP on the horizontal and on the inclined line. The conditions for the stability of solutions  are obtained for the original nonlinear system of equations. For that we use the Lyapunov functions, which are obtained from the conservation laws for the pure proportional feedback controller. We show first integral for the system for the WIP under the proportional feedback controller and we study the areas of the attraction for stability points.

The most part of this work is devoted to studying of the WIP on the soft surface with the rolling friction. It is shown that the stabilization of the WIP under the proportional-derivative controller with the accelerated rotation of the wheel. In the general case such motion leads to the unbounded growth of the rotation velocity of the wheel. The robotic equipment falls due to the limitation of the rotation velocity for the motor on the wheel. We show that the integral term in the proportional-integral-derivative controller allows to stabilize the WIP on two different trajectories. Both of this trajectories tend to the upper position of the pendulum. 

The upper position of the pendulum is semistable. Moreover at this position the wheel states in a stagnation zone due to the rolling resistance on the soft surface.  The physical robotic equipment cannot stay at the semistable position and the WIP changes the direction of the wheel rotation and  falls to the another branch of the movement trajectory.

The same one can see for the numerical calculated trajectories. In this case the transition through the unstable position occurs due to the rounding errors during numeric computations. Such changing between two stable branches of the trajectory looks like the limit cycle. Therefore, we say that the considered proportional-integral-derivative  controller cannot stabilize the WIP  on the soft surface neither  as the physical robotic equipment nor as the numeric simulator.

In section 2 we derive the system of equations for the WIP using the Lagrangian of first kind. In general, the approach is not new but using this approach we remind to the reader the relations  between the  parameters of the dynamical system and  mechanical dimensions of the robotic equipment.

In section 3 we discuss the dynamics of the WIP on the curve without of the control.

In section 4 we consider the stabilization of the WIP on the horizontal. We show that the upper position of the pendulum is asymptotic stable under the proportional-derivative controller. 

In section 5 the dynamics of the WIP on the inclined  line is studied. It is shown  that the  the upper position of the pendulum is stable for the wheel with an unchanged rotating velocity, when the proportional-derivative controller has an additional term with the constant torque.
 
Moving on the soft surface is considered in section 6. The dynamical system for the wheel contains an additional term which defines the friction torque. This additional term has a discontinuous coefficient. It yields the hypersurface in the phase space on which the rotation velocity equals to zero. Such hypersurface is the area of the stagnation of the wheel. In the stagnation area the value of the friction torque is undefined. In this  case the dynamical system  can be modelled by the differential inclusion. The differential inclusions are used in the problem with friction, see for example in the book \cite{Filippov1988Eng}.

As a result the upper point of the pendulum is semistable for the wheel on the soft surface because this point lies in the stagnation region for the torque control  under  the proportional-integral-derivative controller. However in the neighbourhood of the upper point two stable branches of the trajectory appears  which form the sequence of the transitions over the semistable point of the pendulum, and the resulting trajectories of real robotic equipment and the numerical solutions look like limit cycle. This leads to the dynamic stabilization of the pendulum near the upper position.

\section{The system of equations for wheeled inverted Pendulum}

\begin{figure}[h]
\begin{center}
\begin{tikzpicture}[ultra thick]

%Неподвижная платформа
\draw (-1,0)--(1,0);
%\draw (-0.5,0)--(-0.1,0.4);
%\draw (0.5,0)--(0.1,0.4);
%\draw (0,0.5) circle [radius=0.1];

%рычаг маятника
\draw(0,1)--(1,4);
%маятник
\draw[fill=white] (1,4) circle [radius=0.5];
%колесо
\draw(0,1) circle [radius=1];
\draw(0,1) circle [radius=0.1];

%штриховка
\draw (-1,-0.3) --(-0.7,0);
\draw (-0.7,-0.3) --(-0.4,0);
\draw (-0.4,-0.3) --(-0.1,0);
\draw (-0.1,-0.3) --(0.2,0);
\draw (0.2,-0.3) --(0.5,0);
\draw (0.5,-0.3) --(0.8,0);
\draw (0.8,-0.3) --(01.1,0);

%оси координат
\begin{scope}[thin]
\draw (0,0) -- (2,0);
\draw (2.1,0.5) node {$x$} ;
\draw (0,0) --(0,5);
\draw (-0.2,5.1) node {$y$} ;
\draw (0,3) arc[start angle=90,end angle=77,radius=3];
\draw (0.3,3.3) node {$\alpha$} ;
\draw [->](1.3,1) arc[start angle=0,end angle=60,radius=1.3];
\draw (1.3,2) node{$\beta$};
\end{scope}
\end{tikzpicture}
\end{center}
\caption{One can see model of the WIP. Let us define $r$ as the radius of the wheel and $l$ as the length of the pendulum, $\alpha$ is an angle of the pendulum and $\beta$ defines an angle of the turn for the wheel.}
\label{figurePendulumOnTheWheel}
\end{figure}
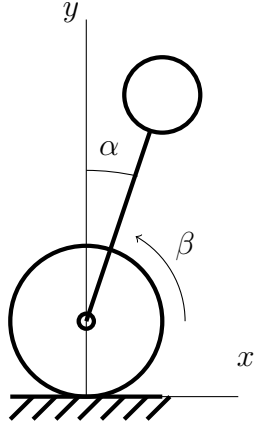
In this section we derive the equations for the dynamics of the inverted pendulum on the wheel which  is presented in the figure \ref{figurePendulumOnTheWheel}. 

Let us define the masses of the pendulum and   the wheel rim as $m$, and $M$. The length of the pendulum equals to $l$ and the radius of the wheel is defined as $r$.   

We will consider the movement along a curve $u(\beta),v(\beta)$, where $u(\beta)$ is the horizontal coordinate and $v(\beta)$ is the vertical coordinate of the point which is the fulcrum of the wheel. In such definitions the kinetic energy is:
$$
T=\frac{m}{2}(\dot{x}^2+\dot{y}^2)+\frac{M}{2}(r^2\dot{\beta}^2+\dot{u}^2+\dot{v}^2).
$$
The potential energy has the form:
$$
\Pi=g v M+g m y.
$$
Using these formulas we can construct the Lagrangian of first kind:
\begin{equation}
L=T-\Pi+\lambda_1((x-u)^2+(y-v)^2-l^2)+\lambda_2(\dot{u}^2+\dot{v}^2-r^2\dot{\beta}^2).
\label{formulaForLagrangeFunction}
\end{equation}

The Lagrangian of the second kind is more conveniently for the study. To derive this Lagrangian we change the coordinates $(x,y)$ and $(u,v)$ to the pair  $\alpha,\beta$. As a result we get:
\begin{eqnarray*}
x=l\sin(\alpha)+u,\quad y=l\cos(\alpha)+v,\\
u=u(\beta),\quad v=v(\beta),\\
\dot{u}=r\dot{\beta}\cos(z),\quad \dot{v}=r\dot{\beta}\sin(z). 
\end{eqnarray*}
Here $z(\beta)$ is the function which defines the inclination of the curve $u(\beta),v(\beta)$.

In such way we obtain the Lagrangian which depends on two independent variables $\alpha,\beta$ and the curve  $z(\beta)$, which defines the position of the wheel. 

The Lagrange equations for the coordinates  $\alpha, \beta$ are the system of two equations of the second order:
$$
\frac{d}{d t}\left(\frac{\partial L}{\partial\dot{\alpha}}\right)-\frac{\partial L}{\partial \alpha}=0,
\quad 
\frac{d}{d t}\left(\frac{\partial L}{\partial\dot{\beta}}\right)-\frac{\partial L}{\partial \beta}=0.
$$
It is convenient to define:
$$
\rho=\frac{r}{l},\quad \zeta=\frac{m}{M},\quad \tau=\sqrt{\frac{g}{l}} t,\quad \frac{d u}{d \beta }=\cos(z),\quad \frac{d v}{d \beta }=\sin(z).
$$

We will define the derivations:
$$
\frac{d f}{d\tau}\equiv\dot{f},\quad \frac{d g}{d\beta}\equiv g'
$$
In such definitions we obtain the system of the equations for $\alpha$ and $\beta$:
\begin{eqnarray}
\ddot{\alpha}=\sin(\alpha)-(\cos(\alpha -z) \ddot{\beta}  +\sin(\alpha -z)z'\dot{\beta}^2 )\rho,
\nonumber\\
(\zeta +2)\rho\,\ddot{\beta}=-\sin(z)-\left(\ddot{\alpha} \cos(\alpha-z)-\dot{\alpha}^2\sin(\alpha-z)\right)  \zeta.
\label{eqForPendulumOnWheelFull}
\end{eqnarray}
Here $z(\beta)$ is the function which defines the curve for rolling  of the wheel.

This system is transformed to the equation for the pendulum if $\rho=0$:
\begin{equation}
\ddot{\alpha}=\sin(\alpha).
\label{EqPendulum}
\end{equation}

If the WIP moves on the horizontal , i.e $z\equiv0$, then we can obtain one equation for the  $\alpha$:
\begin{equation}
(\sin^2(\alpha)\zeta +2)\ddot{\alpha}=(\zeta +2)\sin( \alpha)- \frac{1}{2}\dot{\alpha}^2\zeta\sin(2\alpha).
\label{EqPendulumOnWheel0}
\end{equation}

If the WIP moves on the line with the inclination $z\equiv\varepsilon$, then we can obtain one equation for $\alpha$:
\begin{eqnarray}
(\sin^2(\epsilon-\alpha)\zeta +2)\ddot{\alpha}=
(2+\zeta)\sin(\alpha)  +
\sin(\epsilon)\cos( \epsilon -\alpha)+  
\nonumber\\
\frac{1}{2}\zeta\dot{\alpha}^2 \sin(2(\epsilon -\alpha)). 
\label{EqPendulumOnWheel1}
\end{eqnarray}

The equations (\ref{EqPendulum}) --(\ref{EqPendulumOnWheel1}) have the conservation laws and therefore these equations are integrable. The conservation laws have the forms:
\begin{eqnarray}
E=\frac{\dot{\alpha}^2}{2}+\cos(\alpha),
\\
E_0=\dot{\alpha}^2+(\zeta+2)\cos(\alpha)+
\frac{1}{2}\dot{\alpha}^2\sin^2(\alpha)\zeta,
\\
E_\epsilon=\dot{\alpha}^2+(\zeta+2)\cos(\alpha)+\sin(\epsilon)\sin(\epsilon-\alpha)+
\frac{1}{2}\dot{\alpha}^2\sin^2(\epsilon-\alpha)\zeta.
\end{eqnarray}
Here $E$ is the conservation law for  (\ref{EqPendulum}), $E_0$ is the conservation law for  (\ref{EqPendulumOnWheel0}) and $E_\epsilon$ is the conservation law for (\ref{EqPendulumOnWheel1}).

\section{The stationary points for the wheeled inverted pendulum}
\label{secStationaryPointsForPendulumWithoutControl}

Here we consider the stationary points for the values of $\alpha$, which are solutions of  (\ref{eqForPendulumOnWheelFull}) and  $\alpha\equiv\const$. In particular for the WIP  (\ref{EqPendulumOnWheel0}) the stationary point $\alpha_s=0$ defines the saddle, therefore this point is unstable. For this stationary point the solution for $\beta$ has the form:
$$
\beta=\beta_1\tau+\beta_0,
$$
$\forall \beta_1,\beta_0\in\mathbb{R}$, i.e. the wheel may rotate with a constant velocity.  

For the WIP on the inclined line the equation for the stationary point is:
$$
(2+\zeta)\sin(\alpha_s)  +
\sin(\epsilon)  \cos( \epsilon -\alpha_s)=0.
$$
The solution for this equation has a form:
$$
\alpha_s=\arctan\left(\frac{\sin(2\varepsilon)}{2(2+\zeta+\sin^2(\epsilon))}\right).
$$
In this case the wheel rolls on the inclined line with an acceleration:
$$
\ddot{\beta_s}=\frac{-\sin(\epsilon)}{(\zeta+2)\rho},\quad \hbox{or}\quad 
\beta=\frac{-\sin(\epsilon)}{2(\zeta+2)\rho}\tau^2+\beta_1\tau+\beta_0. 
$$

In the general case the stabilized pendulum leads to override the system of equations for  $\beta(\tau)$ and $\alpha_s=\const$:
\begin{eqnarray}
0=\sin(\alpha_s)-(\cos(\alpha_s -z) \ddot{\beta}  +\sin(\alpha_s -z)z'\dot{\beta}^2 )\rho,
\nonumber\\
(\zeta +2)\rho\,\ddot{\beta}=-\sin(z).
\nonumber
\end{eqnarray}
The condition for the existence of the solution of this overridden system is the equation for the curve $z(\beta)$:
\begin{eqnarray*}
z'\sin(\alpha_s-z)\left(\frac{d}{d\beta}\left(\cos(\alpha_s-z)\frac{\sin(z)}{\zeta+2}\right)+2z'\sin(\alpha_s-z)\frac{\sin(z)}{\zeta+2}\right)=
\\
\left(\sin(\alpha_s)+\cos(\alpha_s-z)\frac{\sin(z)}{\zeta+2}\right)\frac{d}{d\beta}\left(z'\sin(\alpha_s-z)\right).
\end{eqnarray*}

\section{The Stabilization of the wheeled inverted pendulum on the horizontal}
\label{secStabilizationOfThePendulumOnFlatSurface}

The upper stationary point is unstable for the pendulum without an additional control. For the stabilization of the upper point of the pendulum we will rotate the wheel and change the position of the fulcrum for the wheel using the control of the torque on the wheel. 

Let $\mu$ be the torque. If we rewrite this torque  as $\mu=2Mrh$, then the system of the equations for the WIP has a form:
\begin{eqnarray}
\ddot{\alpha}=\sin(\alpha)-(\cos(\alpha -z) \ddot{\beta}  -\sin(\alpha -z)z'\dot{\beta}^2 )\rho-2\frac{\rho}{\zeta} h,
\nonumber\\
(\zeta +2)\rho\,\ddot{\beta}=-\sin(z)-\left(\ddot{\alpha} \cos(\alpha-z)-\dot{\alpha}^2\sin(\alpha-z)\right)  \zeta+\frac{2}{\rho}h.
\label{eqForPendulumOnWheelFullWithStab}
\end{eqnarray}

If the line is the horizontal (i.e. $z\equiv0$), then we get:
\begin{equation}
(\sin^2(\alpha)\zeta +2)\ddot{\alpha}=(\zeta +2)\sin( \alpha)- \frac{1}{2}\dot{\alpha}^2\zeta\sin(2\alpha)-2\left(\frac{1}{\rho}\cos(\alpha)+\left(1+\frac{2}{\zeta}\right)\rho \right)h.
\label{EqPendulumOnWheel0WithStab}
\end{equation}
A typical approach to study the stability and the control of nonlinear dynamical systems is the linear approximation near the stabille solutions. For the  system of equations (\ref{eqForPendulumOnWheelFullWithStab}) and linearization of  (\ref{EqPendulumOnWheel0WithStab}) in the neighbourhood of $\alpha=0$ one can see, for example in \cite{Formalskii2016Eng} (chapter.1 \$2). 

Here we use the approach for the nonlinear system and the conservation law for the WIP under the proportional feedback controller.

Let us consider the proportional-derivative controller:
$$
h=k_1\alpha+k_2\dot{\alpha}.
$$

For $k_2=0$ the equation for the WIP on the horizontal has a following conservation law:
\begin{eqnarray*}
\mathcal{E}_0
&=
\left(1+\frac{1}{2}\sin^2(\alpha)\zeta\right)\dot{\alpha}^2+(\zeta+2)\cos(\alpha)+
\\
&
k_1\left(\frac{2}{\rho}(\alpha\sin(\alpha)+ \cos(\alpha)+\alpha^2\rho\left(1+\frac{2}{\zeta}\right)\right).
\end{eqnarray*}

\begin{figure}
\begin{center}
\includegraphics[scale=0.5]{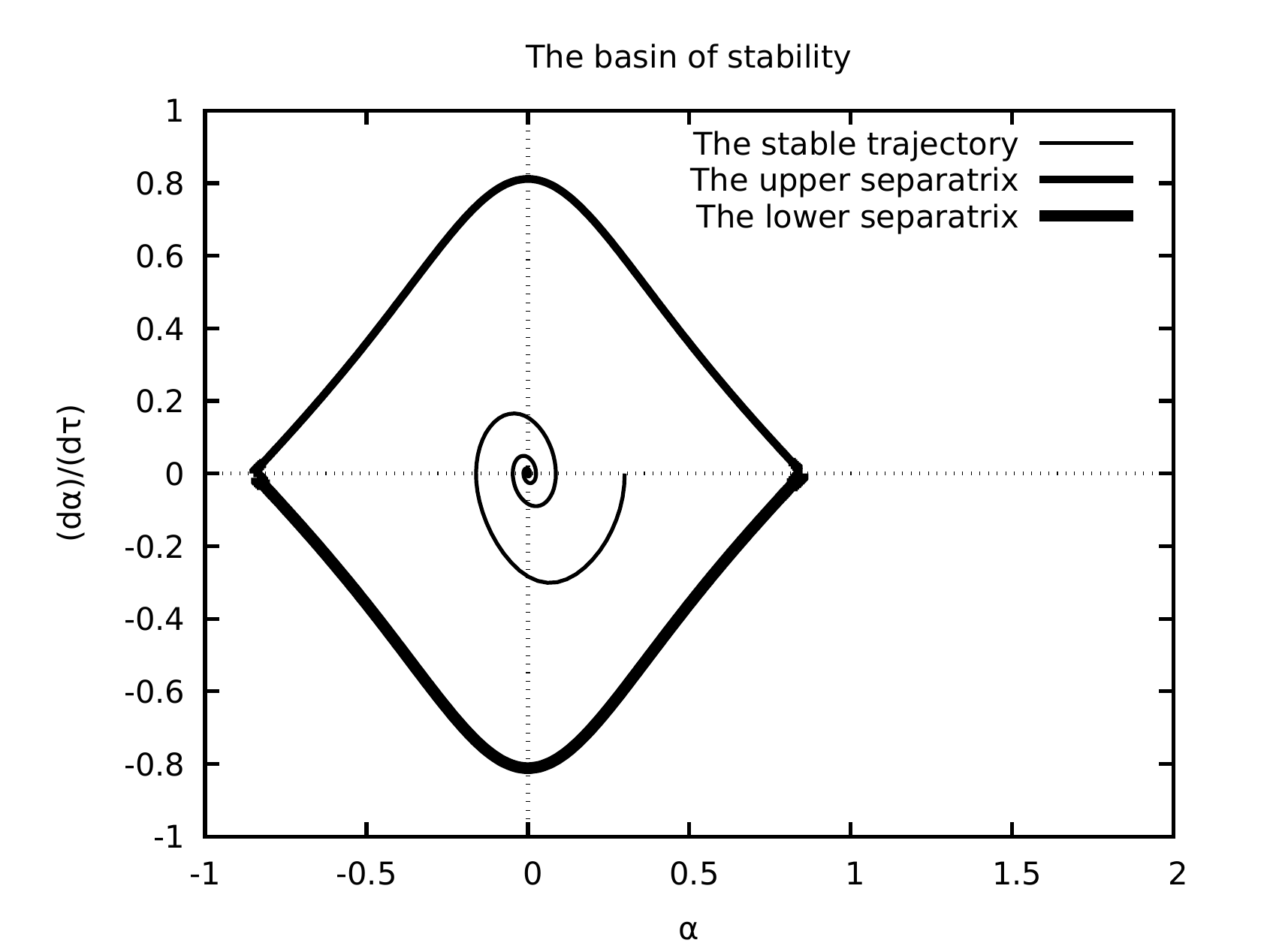}
\end{center}
\caption{Here one can see the phase portrait for the equation  (\ref{EqPendulumOnWheel0WithStab}) in the neighbourhood of the stationary point as $\zeta=10,\,\rho=0.2,\,k_1=1.5,\,k_2=0.05$.  The curves were obtained by numerical solution of  (\ref{EqPendulumOnWheel0WithStab}) by Runge-Kutta forth order.} 
\label{figSeparatrixOnFlatSurface}
\end{figure}

The stationary points  $\alpha_i$ for the pendulum under the control are defined by the equation:
$$
2 k_1 \rho^2(\zeta +2)\alpha -\zeta\rho(\zeta+2)\sin(\alpha) +2k_1\zeta\alpha\cos(\alpha)=0.
$$
If
$$
k_1<\frac{\zeta\rho(2+\zeta)}
{2\zeta+2\rho^2(2+\zeta)},
$$
then there exists the unique stationary point  $\alpha_0=0$, which is a saddle. 

If
\begin{equation}
k_1>\frac{\zeta\rho(2+\zeta)}
{2\zeta+2\rho^2(2+\zeta)},
\label{conditonForPendulumOnHorizontalSurface}
\end{equation}
then there exist three stationary points where  $\alpha_0$ is a stable focus, and if  $k_2=0$, then this point is a center. Two additional points  $\alpha_1>0,\,\alpha_2<0$ are the saddles.

In this case we have two heteroclinic curves, which are the separatrices. One of the separatrices tends to the saddle  $\alpha_1$ as $\tau\to-\infty$ and  to the saddle $\alpha_2$ as  $\tau\to\infty$. The second separatrix tends to the saddle  $\alpha_2$ as $\tau\to-\infty$ and it tends to $\alpha_1$ as $\tau\to\infty$. These separatrices are borders for the oscillatory area near the point  $\alpha_0$, see figure  \ref{figSeparatrixOnFlatSurface}. 

The value of the conservation law on for the separatrices is:
\begin{equation}
\mathcal{E}^{(s)}_0=(\zeta+2)\cos(\alpha_1)+
k_1\left(\frac{2}{\rho}(\alpha_1\sin(\alpha_1)+ \cos(\alpha_1)+\alpha_1^2\rho\left(1+\frac{2}{\zeta}\right)\right).
\label{separatrixForPendulumOnHorizontalSurface}
\end{equation}
 
So the separatrices are the curves on the phase plane $(\alpha,\dot{\alpha})$ with  $\mathcal{E}_s$, see figure \ref{figSeparatrixOnFlatSurface}.

If $k_2>0$, then the derivative of $\mathcal{E}_0$ has a form: 
$$
\dot{\mathcal{E}}_0=-2k_2\dot{\alpha}^2
\left(\left(1+\frac{2}{\zeta}\right)\rho+\frac{1}{\rho}\cos(\alpha)\right).
$$
This formula shows, that the point $(0,0)$ is the stable focus. This point is the attractor for all phase curves in the oscillatory area as $k_2>0$. Second equation of the system (\ref{eqForPendulumOnWheelFull}) shows, that for $(\alpha,\dot{\alpha})=(0,0)$  one gets $\ddot{\beta}=0$, i.e. the wheel can rotate with a constant velocity.

The result of this section is following.
\begin{theorem}
\label{theoremAboutStabPDOnHorizontal}
 If (\ref{conditonForPendulumOnHorizontalSurface}), $k_2>0$ and the trajectories of the equation (\ref{EqPendulumOnWheel0})  lie into the area between the separatrices which are defined by the equation  (\ref{separatrixForPendulumOnHorizontalSurface}), then such trajectories tend to $(0,0)$ as $t\to\infty$. The acceleration of the wheel $\ddot{\beta}=0$ at the point $(\alpha,\dot{\alpha})=(0,0)$.
\end{theorem}
 
\section{The stabilization on the inclined line}
\label{secStabilizationOfThePendulumOnSlantedSurface}

The equation for the WIP on the inclined line as $z\equiv\epsilon=\const$ looks following:
\begin{eqnarray}
(\sin^2(\epsilon-\alpha)\zeta +2)\ddot{\alpha}=
(2+\zeta)\sin(\alpha)  +
\sin(\epsilon)  \cos( \epsilon -\alpha) +
\nonumber\\
\frac{1}{2}\zeta\dot{\alpha}^2 \sin(2(\epsilon -\alpha))-
2\left(
\frac{1}{\rho}\cos(\epsilon-\alpha)+
\left(1+\frac{2}{\zeta}\right)\rho
\right) h. 
\label{EqPendulumOnWheel1WithStab}
\end{eqnarray}

In the section  \ref{secStationaryPointsForPendulumWithoutControl} it is shown, that without of the control the wheel should be accelerate for keeping of the pendulum in the upper position. Such behaviour is not appropriated for the robotic equipment. Therefore we use the controller to stabilize the WIP.

The problem about the stabilization of the inverted pendulum was considered for linearized equations on the inclined surface for example in  \cite{SamiMichalskaAngeles2007}, see also  \cite{PathakFranchAgrawal2005}. Here we study the more wide problem for the pendulum including the area, where  the nonlinearity is important.

Let the control be following:
\begin{equation}
h=k_1\alpha+k_2\dot{\alpha}.
\label{formulaForControlOfPendulum1}
\end{equation}

The equation for $\alpha$ has the form:
\begin{eqnarray}
(\sin^2(\epsilon-\alpha)\zeta +2)\ddot{\alpha}=
(2+\zeta)\sin(\alpha)  -\frac{1}{2}\zeta\dot{\alpha}^2 \sin(2(\epsilon -\alpha))-  
\nonumber\\
 2\rho\cos(\epsilon-\alpha)(k_1\alpha+k_2\dot{\alpha}). 
\label{EqPendulumOnWheel1WithStab1}
\end{eqnarray}

If $k_2=0$, then the equation (\ref{EqPendulumOnWheel1WithStab1}) has the conservation law:
\begin{eqnarray}
\mathcal{E}_\epsilon=
& k_1 \left(\rho\left(1+\frac{2}{\zeta}\right)\alpha^2+\frac{2}{\rho}(\cos(\epsilon-\alpha)-2\alpha\sin(\epsilon-\alpha))\right) 
 +
\nonumber\\
&
\cos(\alpha)(\zeta +2)
+\sin(\epsilon)\sin(\epsilon-\alpha)+\left(\frac{1}{2}
\,\sin^2(\epsilon-\alpha)\zeta+1\right)\dot{\alpha}^2.
\label{formulaForConservationLawOnSlantedSurface}
\end{eqnarray}
In this case the equation is integrable.

The stationary points for   (\ref{EqPendulumOnWheel1WithStab1}) are the solutions of the equation:
\begin{eqnarray}
2((\zeta +2)\rho^2+ 
\zeta\cos(\epsilon-\alpha))k_1 \alpha
&
+
&
\\
((-\sin(\epsilon)\cos(\epsilon-\alpha) -2 \sin(\alpha))\zeta -\sin(\alpha)\zeta^2)\rho 
&
=
&
0.
\label{EqForEquilibriumPointsOnSlanetedLine}
\end{eqnarray}
\begin{figure}
\includegraphics[scale=0.5]{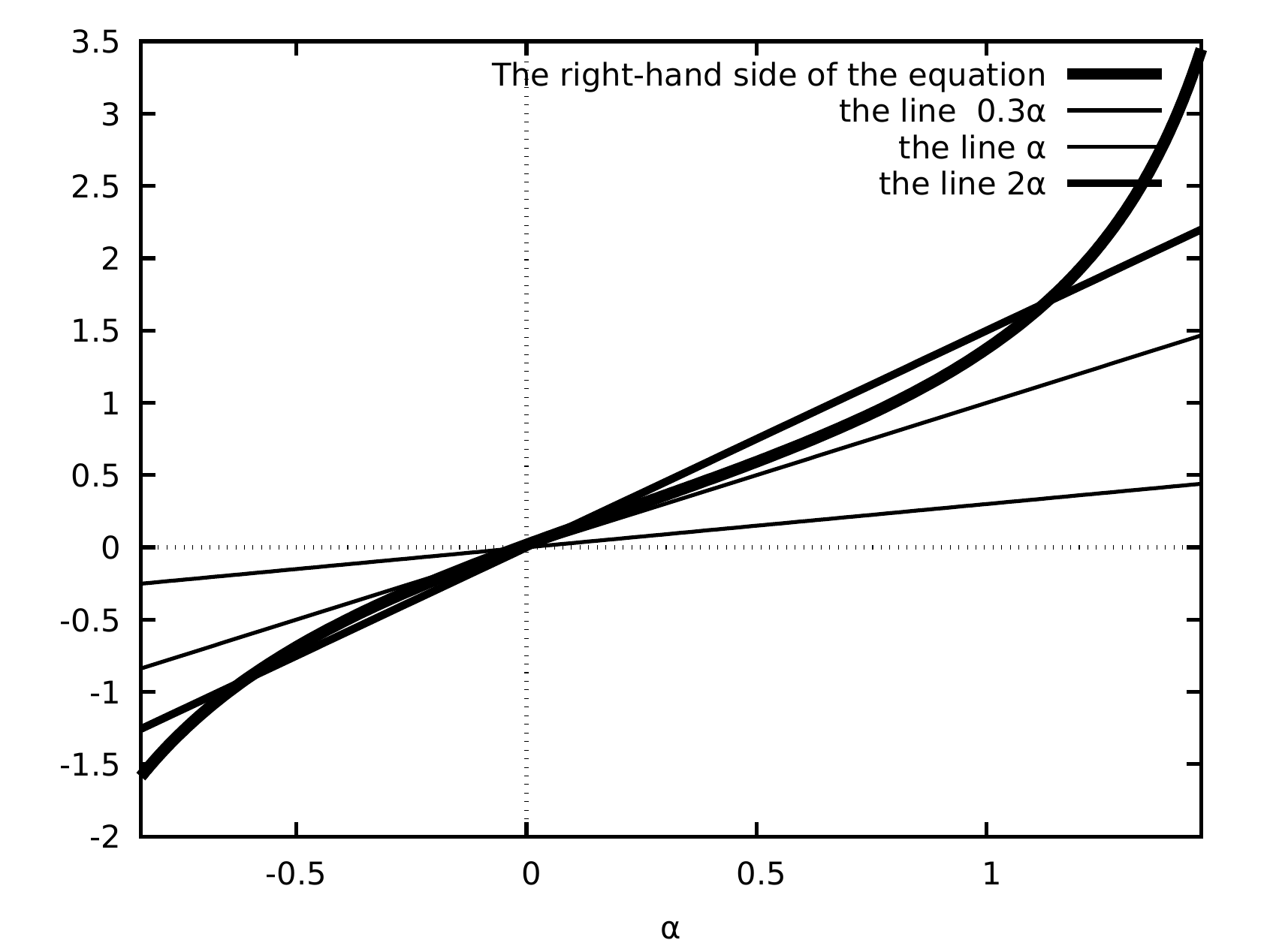}
\caption{The solutions of the equation (\ref{EqForStabPointsOnSlantedLine}) are the cross points between thick curve and the straight lines. Here  $\rho=0.2$, $\zeta=10$, $\epsilon=0.2$}
\label{FigForConstructionOfStabPointsOnSlantedLine}
\end{figure}

The solutions of (\ref{EqForEquilibriumPointsOnSlanetedLine})  cannot be write in the obviously form. We rewrite this equation into the more convenient form for the analysis:
\begin{eqnarray}
k_1\alpha=\frac{(\sin(\alpha)(\zeta +2) +\sin(\epsilon) \cos(\epsilon-\alpha))\zeta  \rho}
{2(\zeta+2)\rho^2 +2\cos(\epsilon-\alpha)\zeta}.
\label{EqForStabPointsOnSlantedLine}
\end{eqnarray}
The graphical solutions of this equation are shown in  figure \ref{FigForConstructionOfStabPointsOnSlantedLine}. If $k_1=2$, then there exist three crossing points. But for small values of $k_1$, such as $k_1=0.3$ there exists one such point only.

It can be shown that $\exists\epsilon_0$, $0\le\epsilon_0<\pi/2$, such that for $\epsilon\in[0,\epsilon_0)$, there exist $k_1$, such that the equation for $\alpha\in(-\pi/2,\pi/2)$ has three stationary solutions $\alpha_{0,1,2}$, such that $-\pi/2\alpha_1<\alpha_0<\alpha_2<\pi/2$.

The points  $\alpha_{1,2}$ are the saddles on the phase plane $(\alpha,\dot{\alpha})$. The point $\alpha_0$ is the center as $k_2>0$.

\begin{figure}
\includegraphics[scale=0.5]{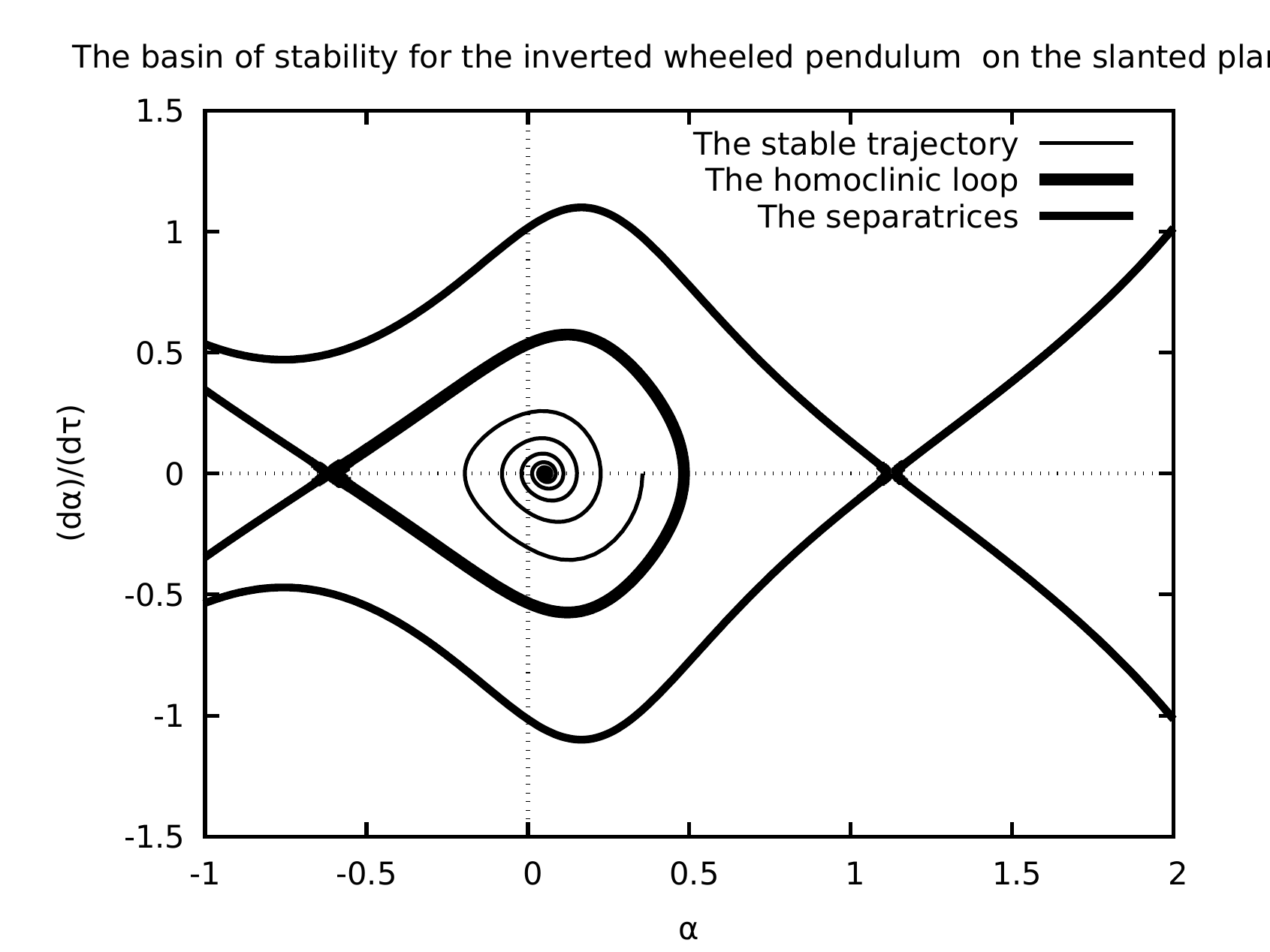}
\caption{On this figure one can see the homoclinics which are corresponding to the left and right saddle points. The homoclinic loop is a border for trajectories of the pendulum which tend to $(\alpha_0,0)$. These curve were constructed as the numerical solutions of  (\ref{EqPendulumOnWheel1WithStab})  as $\zeta=10,\,\rho=0.2,\,k_1=1.5$. The trajectory, which tends to  $(\alpha_0,0)$, has a parameter $k_2=0.05$.}
\label{figSeparatrixForSlantSurface}
\end{figure}

The values of the conservation law $\mathcal{E}_\epsilon$ are differ in the saddle points in general. Therefore there exist two homoclinic trajectory  for each of the saddles such that the trajectories tend  to $\alpha_i$ as $t\to\pm\infty$ and  $k_2=0$. The region of the stability for the pendulum is defined by smallest of the homoclinic loops, see figure \ref{figSeparatrixForSlantSurface}. 

Let us differentiate the function $\mathcal{E}_\epsilon$ using (\ref{EqPendulumOnWheel1WithStab}) as $k_2>0$. As a result we get:
$$
\frac{d\mathcal{E}_\epsilon}{d\tau}=
-2k_2\dot{\alpha}^2f_\epsilon(\alpha,\zeta,\rho),
$$
where
$$
f_\epsilon(\alpha,\zeta,\rho)=\left(
\left(1+\frac{2}{\zeta}\right)\rho+
\frac{1}{\rho}\cos(\epsilon-\alpha)
\right).
$$
If 
\begin{eqnarray}
f_\epsilon(\alpha,\zeta,\rho)>0,\quad \forall\alpha\in(\alpha_1,\alpha_m), \quad
\alpha_m=\max_{\alpha\in\mathcal{L}} (\alpha), \quad  
\mathcal{L}=\{(\alpha,\dot{\alpha})|  \mathcal{E}_\epsilon|_{\alpha=\alpha_1}\},
\label{conditionOfStability1}
\end{eqnarray}
then all trajectories tend to smallest value of 
$\mathcal{E}_\epsilon$, i.e. the trajectories tend to $(\alpha_0,0)$.

Let us formulate the result of this section:

\begin{theorem}
\label{theoremAboutStabPDOnSlantedLine}
The phase portrait of the equation (\ref{EqPendulumOnWheel1WithStab}) has the asymptotic stable point $(\alpha_0,0)$ for  $h$ such as  (\ref{formulaForControlOfPendulum1}) with $k_1$ and $k_2$ such as  (\ref{conditionOfStability1}).
The trajectories within the homoclinic loop  (\ref{figSeparatrixForSlantSurface}) as  $\alpha=\alpha_i$, where 
$|\alpha_i|<|\alpha_j|,\,i,j=1,2$ tend to $(\alpha_0,0)$ when $\tau\to\infty$.  In the point $(\alpha,\dot{\alpha})=(\alpha_0,0,)$ the acceleration of the wheel equals to zero.
\end{theorem}

\section{The stabilization on the wheeled inverted pendulum on soft surface}
\label{secStabilizationOfThePendulumOnSoftSurface}

When the wheel is moving on the horizontal soft surface the wheel lies into the small hole. Therefore it needs an non-zero torque to begin the moving.
 
The movement on the horizontal soft surface is defined by the differential inclusion:
\begin{eqnarray}
\ddot{\alpha}-\sin(\alpha)+2\frac{\rho}{\zeta}h = \rho \cos(\alpha)\ddot{\beta},
\nonumber\\
(\zeta +2)\rho\,\ddot{\beta}\in 
F(\alpha,\dot{\alpha},\ddot{\alpha},\dot{\beta}).
\label{eqForPendulumOnWheelFullWithStabAndDump}
\end{eqnarray}
Let us define:
$$
f=-\left(\ddot{\alpha}\cos(\alpha) -\dot{\alpha}^2\sin(\alpha)\right)  \zeta+\frac{2}{\rho}h.
$$ 
The map  $F(\alpha,\dot{\alpha},\ddot{\alpha},\dot{\beta})$ has a form:
$$
F(\alpha,\dot{\alpha},\ddot{\alpha},\dot{\beta})
=
\left\{
\begin{array}{cc}
f-\nu\ \sgn(\dot{\beta}),
&
\quad\quad \{\forall\dot{\beta}\not=0\};
\\
(-\nu,\nu),
&
\quad\{\dot{\beta}=0\}\cup\{ |f|\le\nu\};
\\
f-\nu\ \sgn(\dot{\beta}),
&
\quad \{\dot{\beta}=0\}\cup\{\{\alpha,\dot{\alpha},\ddot{\alpha}\} \in \{|f|>\nu\}\}. 
\end{array}
\right.
$$

It is convenient to consider differential inclusion  (\ref{eqForPendulumOnWheelFullWithStabAndDump}) as two dynamical systems. One of the system defines the blocked wheel and another one system defines the rotating wheel of the WIP.

If $\dot{\beta}\equiv0$, then the equation for the pendulum is following:
\begin{equation}
\ddot{\alpha}-\sin(\alpha)+\frac{\rho}{\zeta}(k_1\alpha+k_2\dot{\alpha})=0.
\label{EqForPendulumWith BlockedWheel}
\end{equation}
This is the equation for the nonlinear pendulum with a dissipation and and additional torque. 

The equilibrium point for (\ref{EqForPendulumWith BlockedWheel}) is following: $(\alpha,\dot{\alpha})=(0,0)$. If $k_1>(\zeta/\rho)$, then this equilibrium is the  unique stable stationary point. 

If $k_1<(\zeta/\rho)$, the solution $(\alpha,\dot{\alpha})=(0,0)$ is unstable, and for $\alpha\in(-\pi,\pi)$ there exist two additional stationary solutions of this equation, such that $-\pi<\alpha_1<0<\alpha_2<\pi$. These solutions are  the stable focuses as $k_2>0$. 

If $k_1< (\zeta/\rho)$ and $2k_1\rho|\alpha_{1,2}|\le\nu$, then there exist two stable stationary positions of the pendulum for the blocked wheel.

If $k_1< (\zeta/\rho)$, and $2k_1\rho|\alpha_{1,2}|>\nu$, then there is no a stable equilibrium for the pendulum with the blocked wheel.  

Let us consider the dynamics of the WIP on the branches of the trajectory with the condition $\sgn(\dot{\beta})=\const$. The equation for $\alpha$ is following:
\begin{eqnarray*}
(2+\zeta\sin^2(\alpha))\ddot{\alpha}=
&
(2+\zeta)\sin(\alpha)-\frac{\zeta}{2}\dot{\alpha}^2\sin(2\alpha)-2\rho\left(1+\frac{2}{\zeta}\right) (k_1\alpha+k_2\dot{\alpha})-
\\
&
\frac{2}{\rho}\cos(\alpha)(k_1\alpha+k_2\dot{\alpha})+\sgn(\dot{\beta})\nu\cos(\alpha).
\end{eqnarray*}

If $\sgn(\dot{\beta})=\const$ and $k_2=0$, then this equation has the conservation law:
\begin{eqnarray*}
\mathcal{E}_\nu
&=
\left(1+\frac{1}{2}\sin^2(\alpha)\zeta\right)\dot{\alpha}^2+(\zeta+2)\cos(\alpha)
-\nu\sgn(\dot{\beta})\cos(\alpha)
+
\\
&
k_1\left(\frac{2}{\rho}(\alpha\sin(\alpha)+ \cos(\alpha))+\alpha^2\rho\left(1+\frac{2}{\zeta}\right)\right).
\end{eqnarray*}

If $k_2\not=0$, then:
$$
\dot{\mathcal{E}}_\nu=-2k_2\dot{\alpha}^2
\left(\left(1+\frac{2}{\zeta}\right)\rho+\frac{1}{\rho}\cos(\alpha)\right).
$$

If $\dot{\beta}\not=0$, then we obtain the equation for the stationary points:
\begin{eqnarray}
\frac{k_1\rho}{\zeta}\alpha=
\frac{(\nu\,\sgn(\dot{\beta})\cos(\alpha)+(\zeta+2)\sin(\alpha))
\rho^2}
{2(\zeta+2)\rho^2+2\cos(\alpha)\zeta}.
\label{EqForStationaryPointsOnSoftLineForPD}
\end{eqnarray}
\begin{figure}
\includegraphics[scale=0.5]{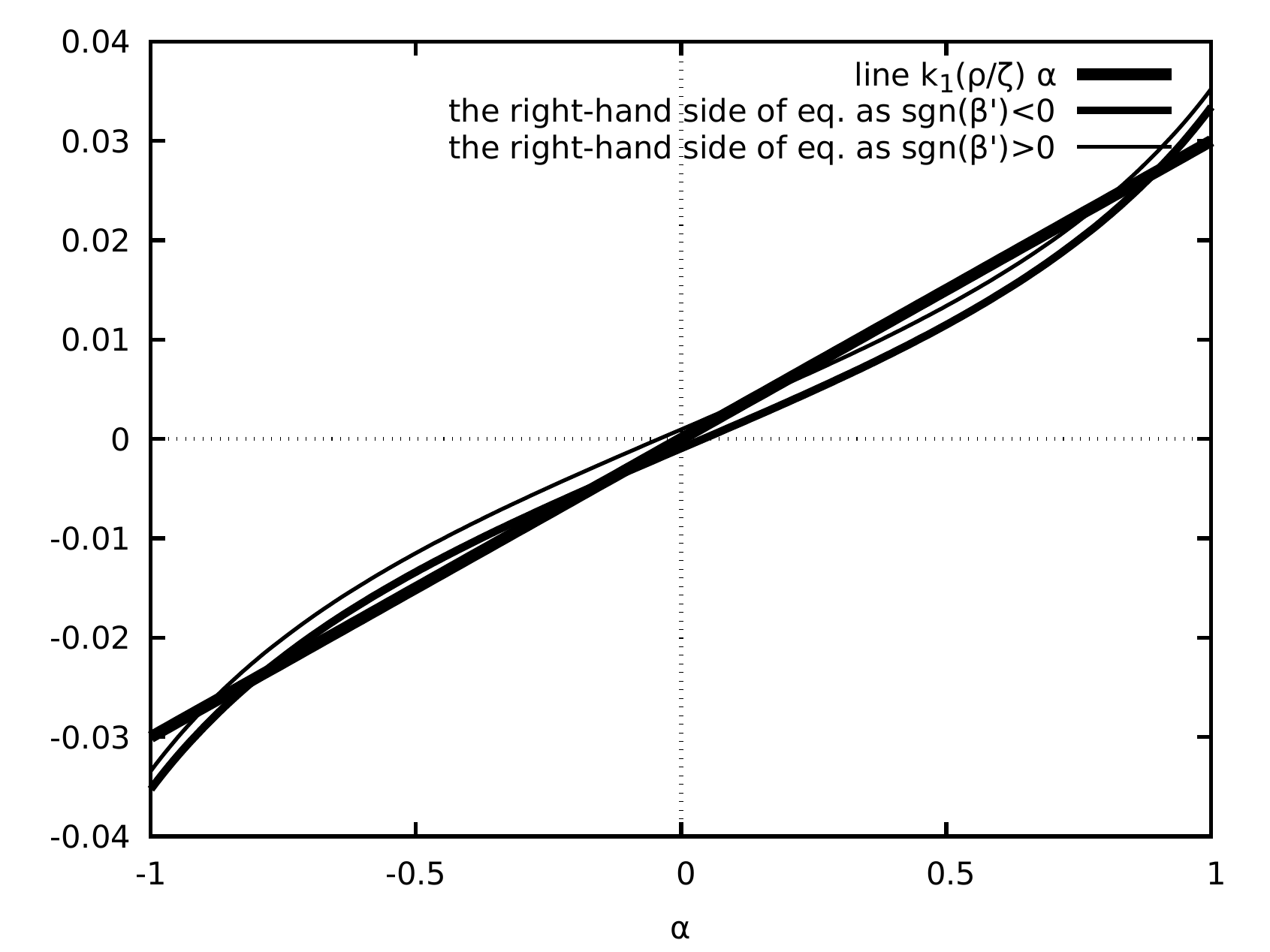}
\caption{
Here two curves correspond right-hand side of the equation  (\ref{EqForStationaryPointsOnSoftLineForPD}) as $\sgn(\dot{\beta})>0$ and $\zeta=10$, $\rho=0.2$, $\nu=0.5$. The straight line corresponds to the left-hand side of the  equation (\ref{EqForStationaryPointsOnSoftLineForPD}).}
\label{figStationaryPointsOnSoftSurfaceForPD}
\end{figure}
This equation has three solutions as  $\sgn(\dot{\beta})=\pm1$ $-\pi/2<\alpha_1^\pm<\alpha_0^\pm<\alpha_2^\pm<\pi/2$, then  the function  $\mathcal{E}_\nu$ has a minimum at the point 
$\alpha_0^\pm$. In these cases $\sgn(\dot{\beta})=\pm1$  the points of the minima of $\mathcal{E}_\nu$ are such, that  $\alpha_0^-<0<\alpha_0^+$.

\begin{figure}
\includegraphics[scale=0.3]{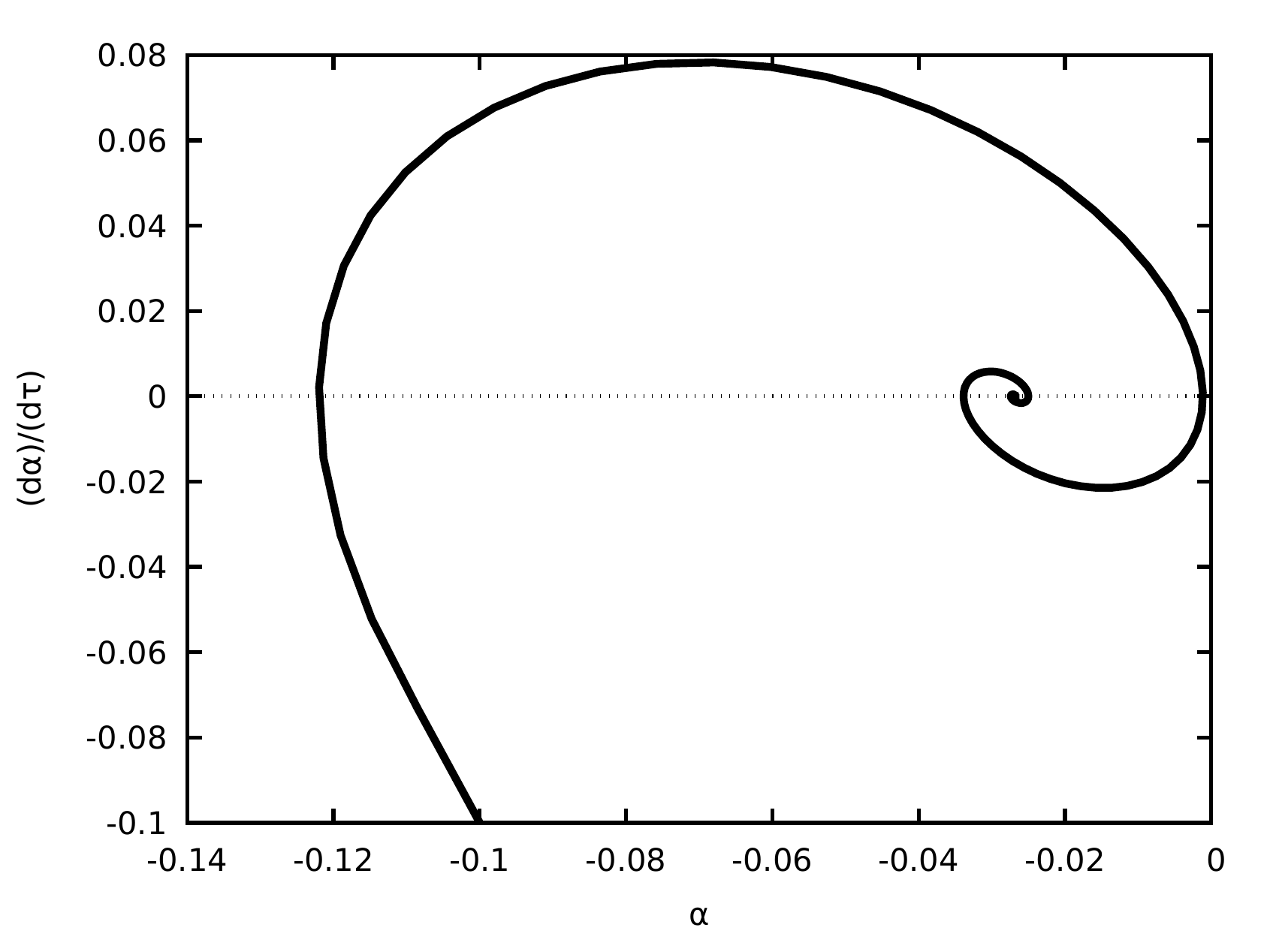}
\includegraphics[scale=0.3]{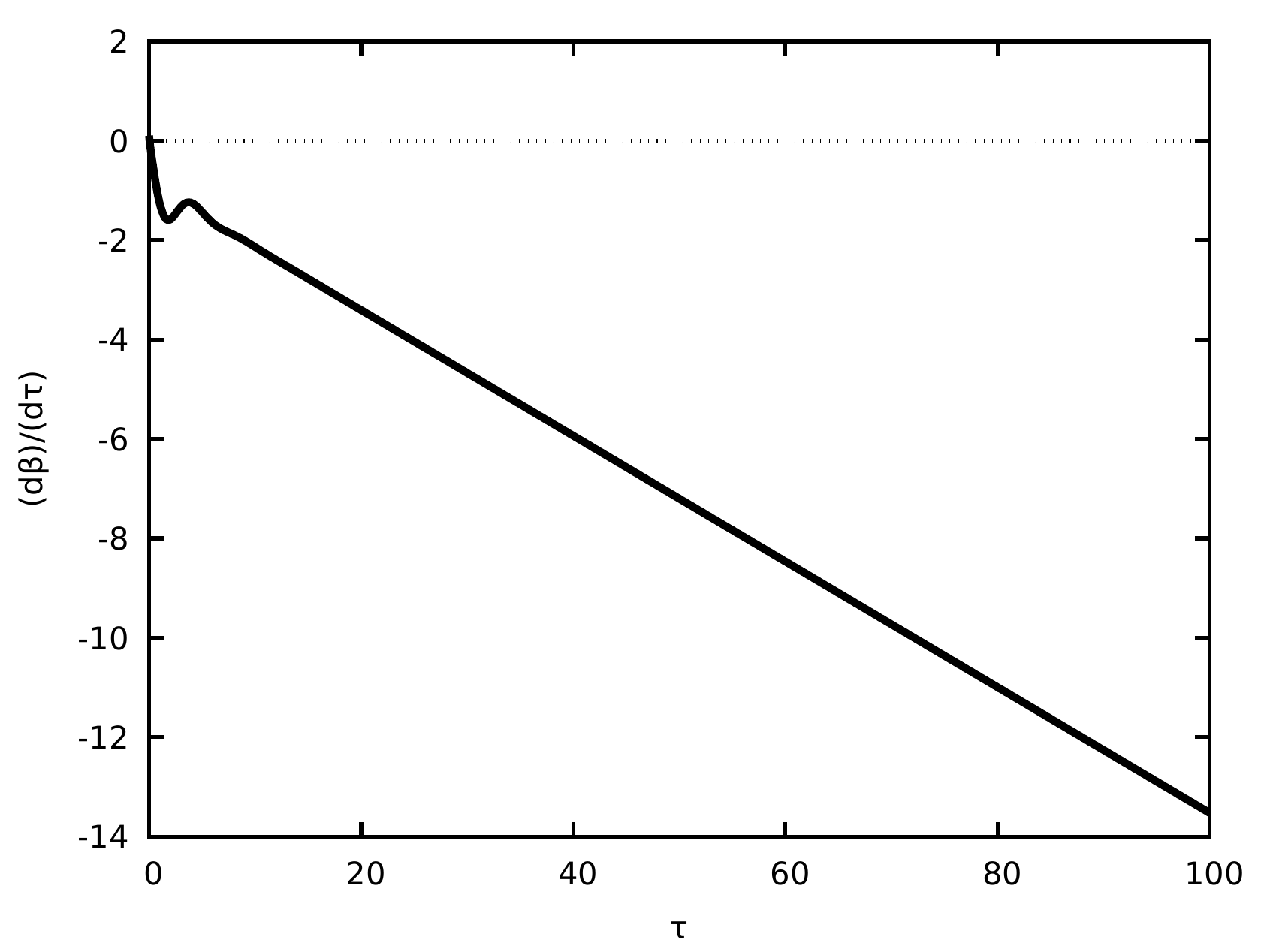}
\caption{
Here one can see the solutions of the equations (\ref{eqForPendulumOnWheelFullWithStabAndDump}) with the  initial conditions $\tau=0$, $\alpha=-0.1,\dot{\alpha}=-0.1,\beta=0,\dot{\beta}=0.1$ and  the proportional-derivative controller, where $k_1=1.5,k_2=0.2$. On the left-hand side   one can see the stabilization of $\alpha$, on the right-hand side one can see the linear dependency of  $\beta(\tau)$ for large $\tau$. Here the coefficient of the friction is $\nu=0.05$.
}
\label{figRotationOfWheelOnSoftPlaneUnderPDController}
\end{figure}

The stabilization of the pendulum in the points  $\alpha_0^\pm$ as $2k_2|\alpha_0|>\rho\nu$ leads to  acceleration of the wheel:
\begin{eqnarray*}
\ddot{\beta_{\pm}}=\frac{1}{\rho(\zeta+2)}\left(
\frac{2}{\rho}k_1\alpha_0^{\pm}-\nu\sgn(\dot{\beta})
\right).
\end{eqnarray*}  

The permanently acceleration of the wheel is not appropriate behaviour of the solutions for robotics equipment. The numerical example of such solution is shown in the figure  
\ref{figRotationOfWheelOnSoftPlaneUnderPDController}.   

If $2k_2|\alpha_0^\pm|\le\rho\nu$, then $\ddot{\beta}=0$ and for $\alpha_0^\pm$ one get the following inequality:
\begin{eqnarray}
-\nu<\frac{\zeta+2}{\cos(\alpha_0^\pm)}
\left(-\sin(\alpha_0^\pm)+2\frac{\rho}{\zeta}k_1\alpha_0^\pm \right)<\nu.
\end{eqnarray}

Thus to stabilize the WIP one should find the control, which stabilizes the pendulum in the point  $\alpha=0$ or in the small neighbourhood of this point.

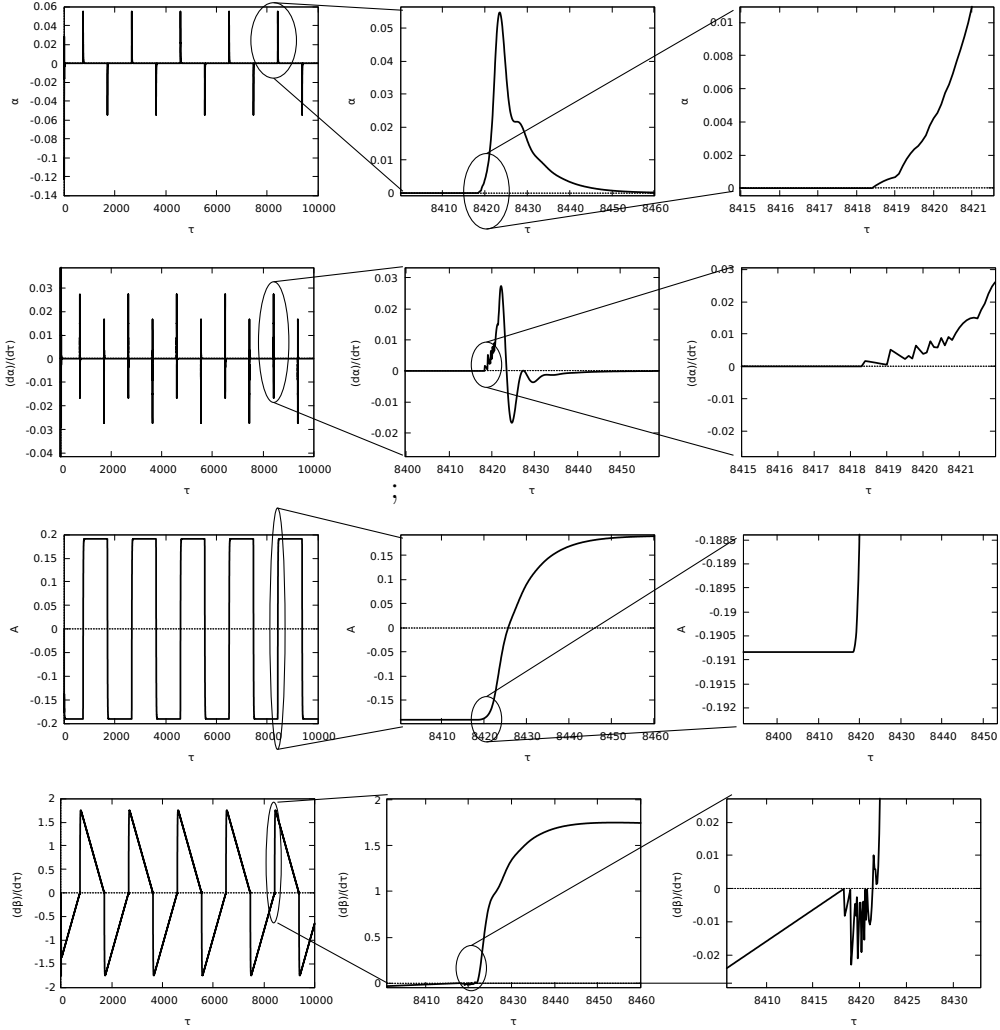
\begin{figure}
\begin{tikzpicture}[thin]

\node at (-3,1.5)
{
\hspace{-1cm}
\includegraphics[scale=0.33]{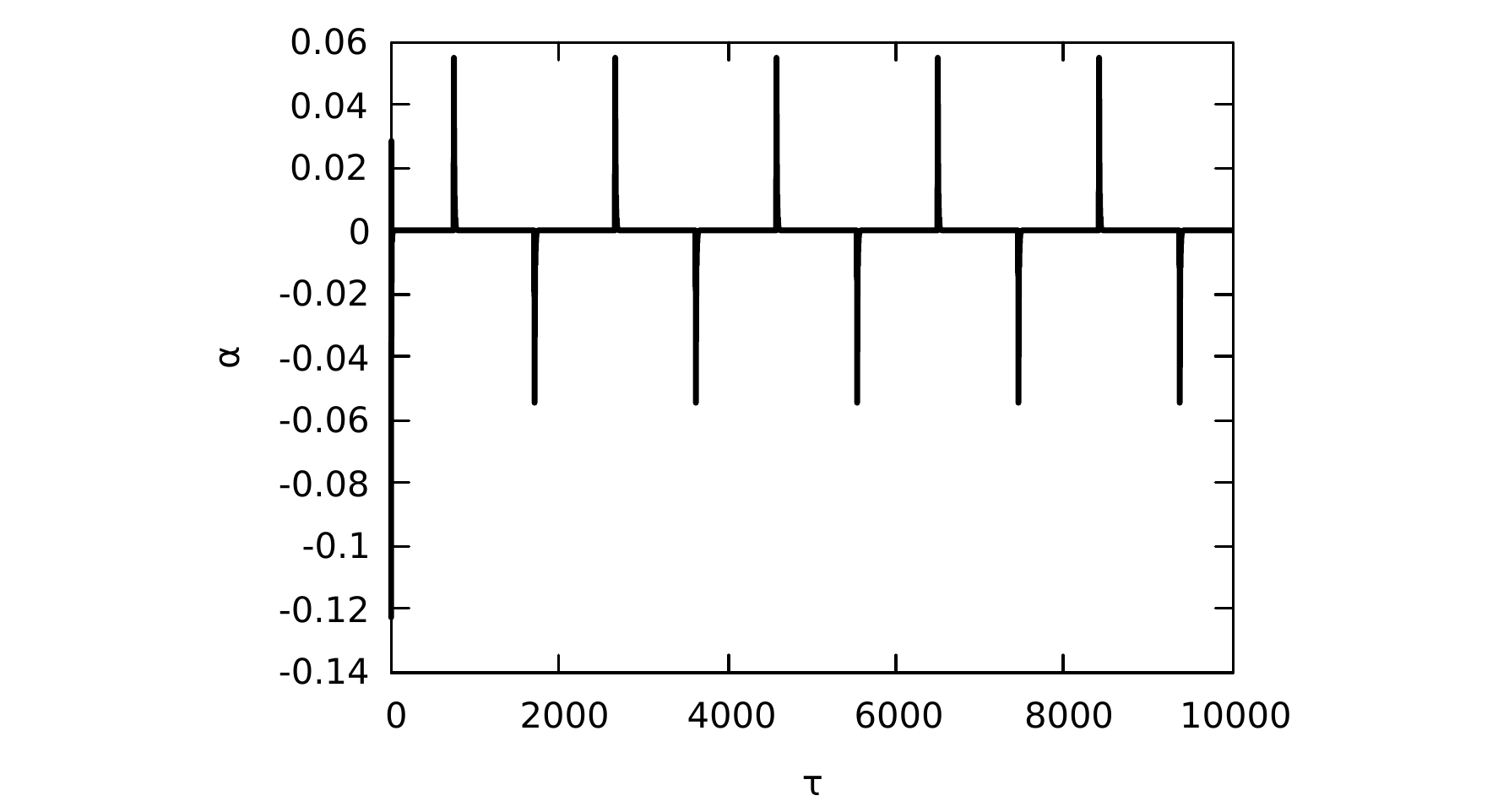}
\hspace{-1.75cm}
\includegraphics[scale=0.33]{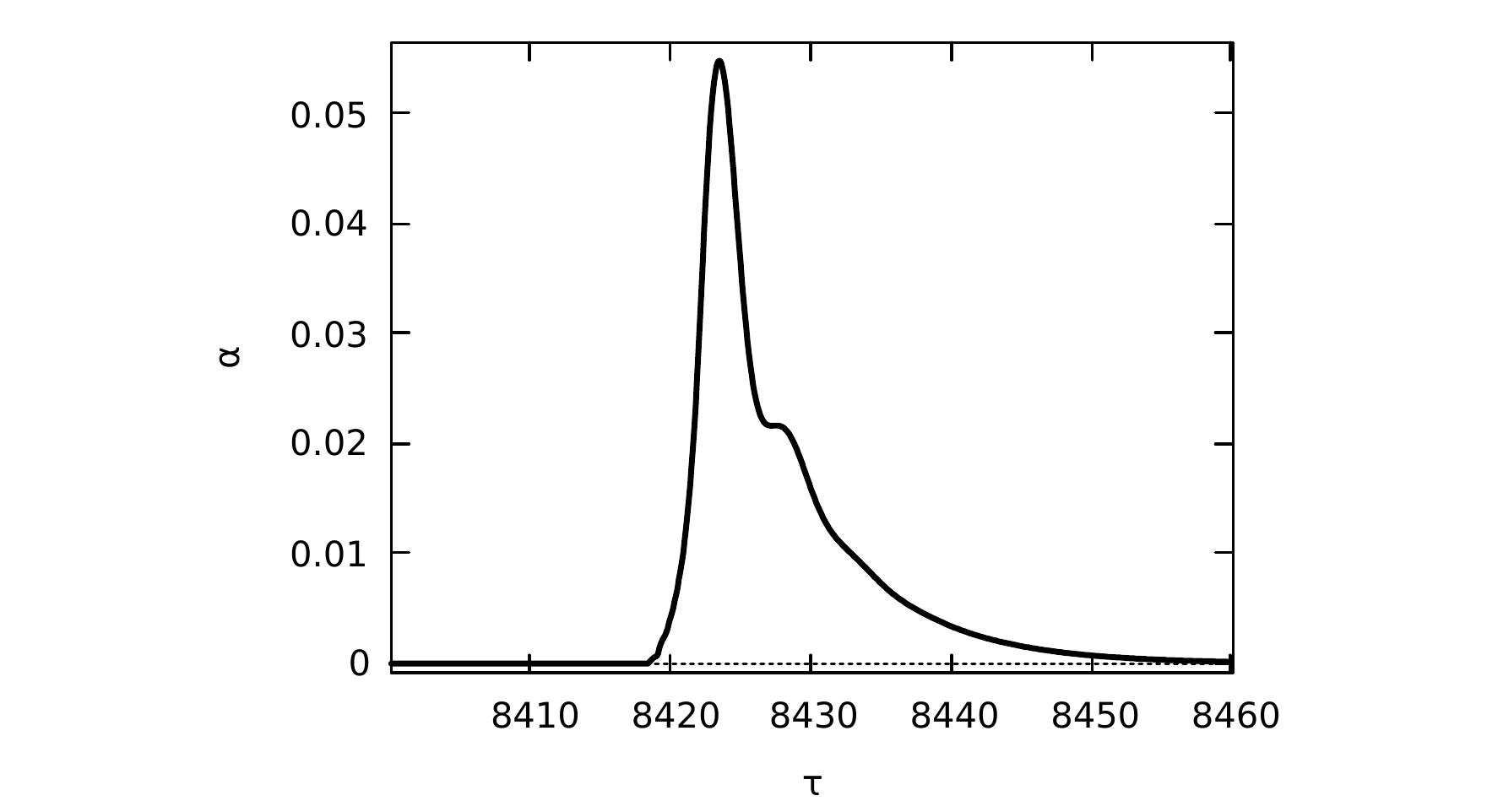}
\hspace{-1.75cm}
\includegraphics[scale=0.33]{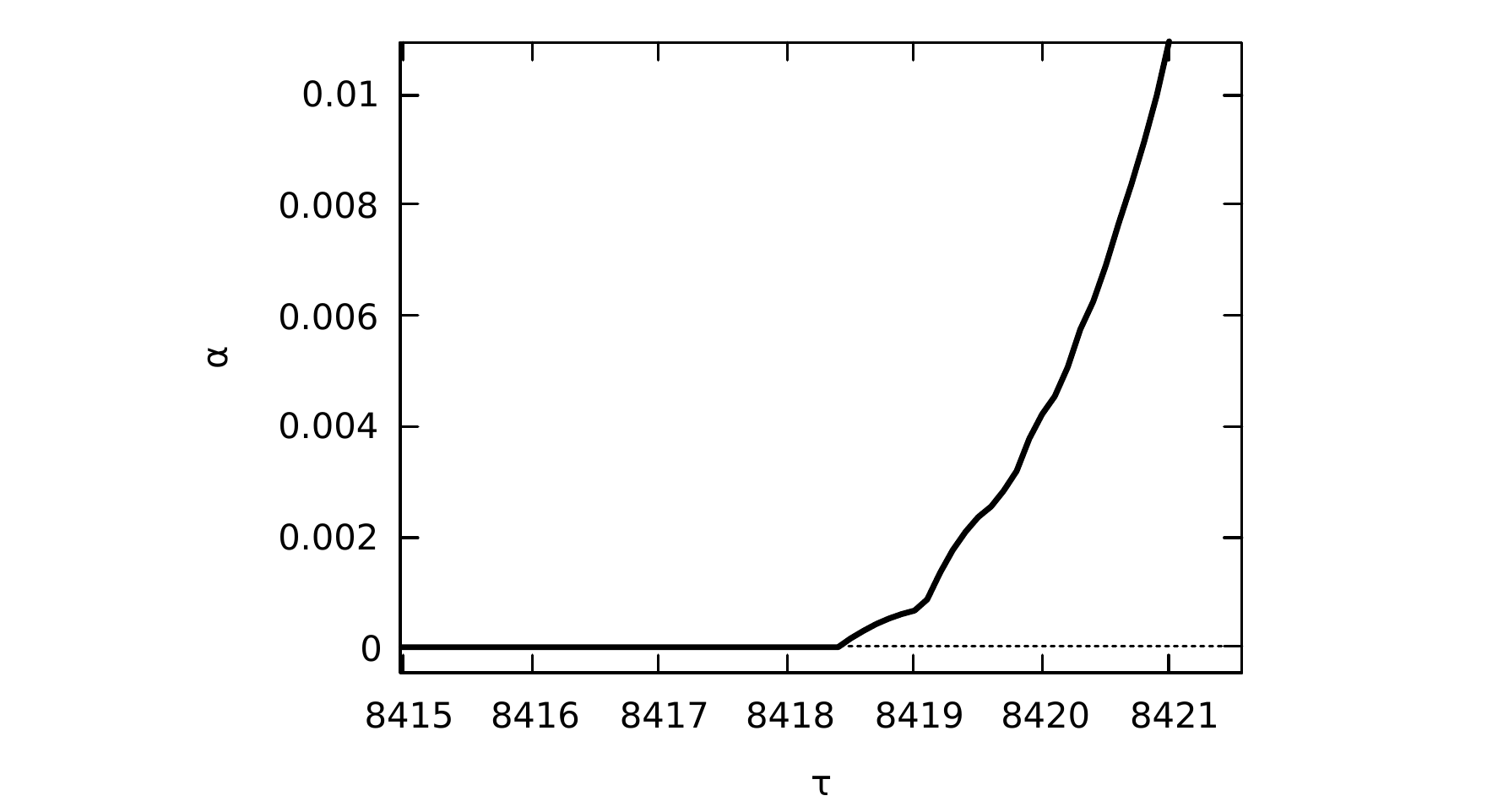}
};
\draw(-6.5,2.5) ellipse (0.3cm and 0.5cm);
\draw (-6.5,3)--(-4.8,2.9);
\draw (-6.5,2)--(-4.8,0.5);

\draw(-3.7,0.5) ellipse (0.3cm and 0.5cm);
\draw (-3.7,1)--(-0.45,2.9);
\draw (-3.7,0)--(-0.45,0.5);

\node at (-3,-2) 
{
\hspace{-1cm}
\includegraphics[scale=0.33]{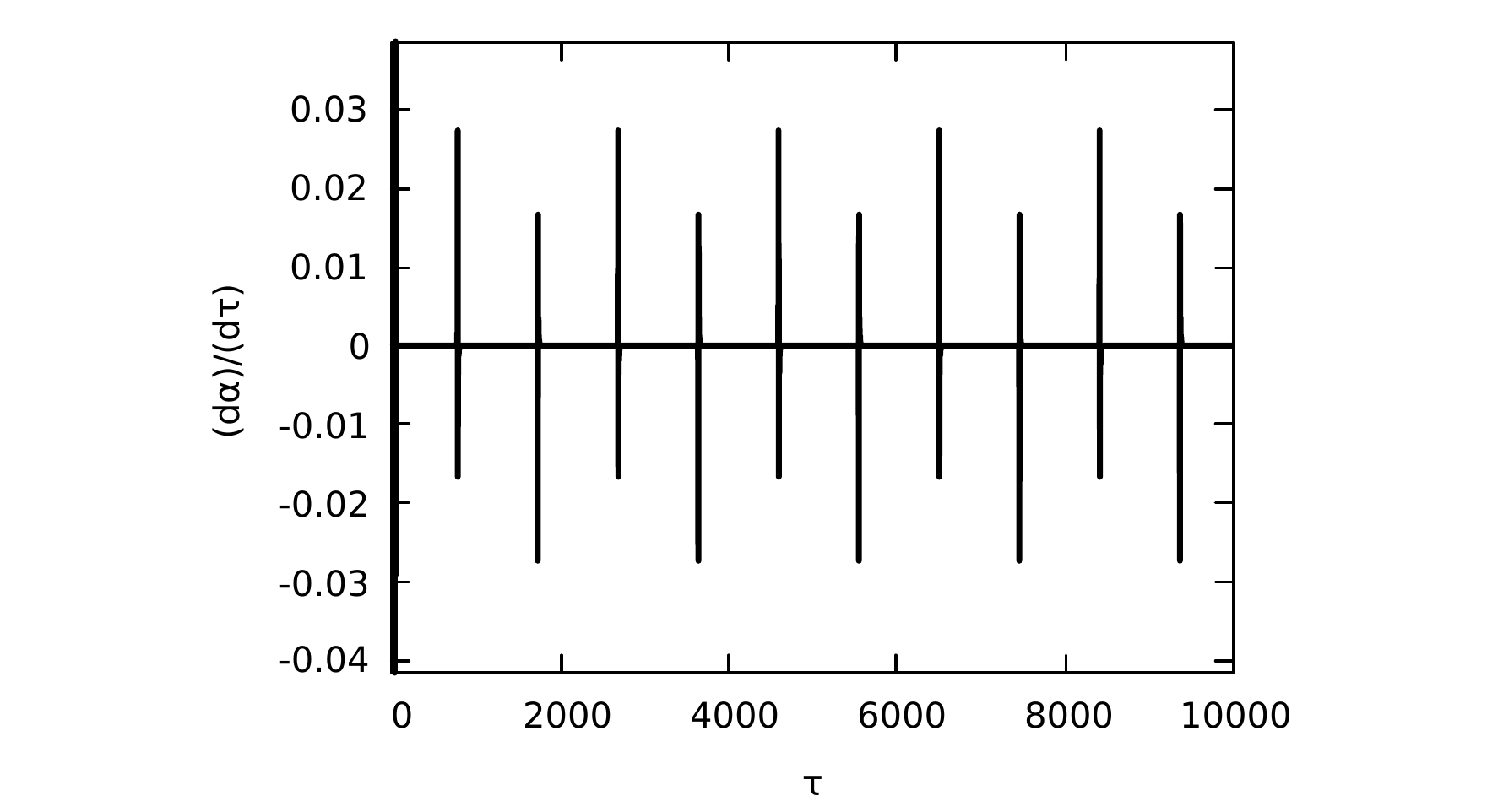};
\hspace{-1.75cm}
\includegraphics[scale=0.33]{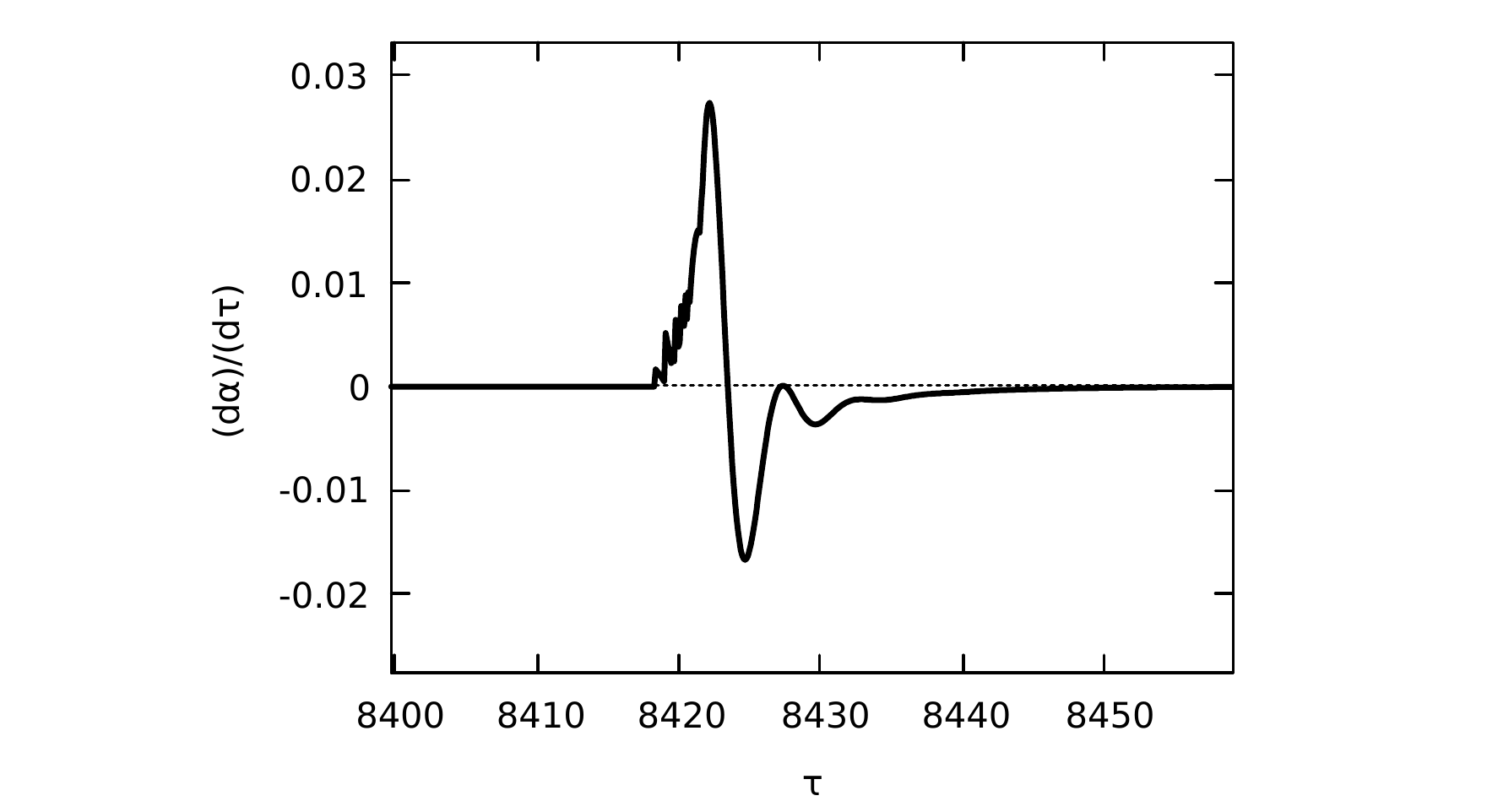}
\hspace{-1.75cm}
\includegraphics[scale=0.33]{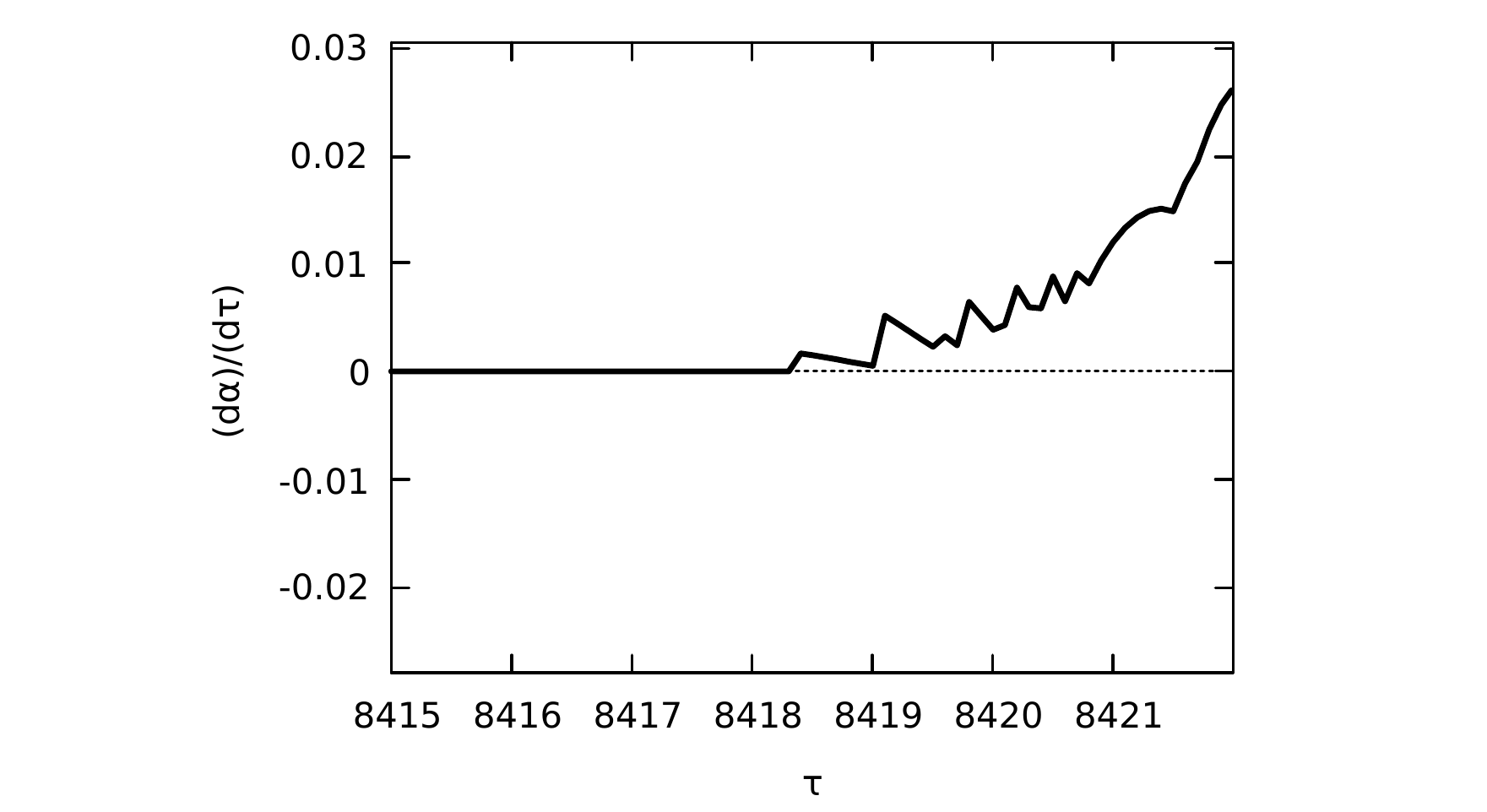}
};

\draw(-6.5,-1.5) ellipse (0.2cm and 0.8cm);
\draw (-6.5,-0.7)--(-4.8,-0.5);
\draw (-6.5,-2.3)--(-4.8,-3);

\draw(-3.7,-1.8) ellipse (0.2cm and 0.3cm);
\draw (-3.7,-1.5)--(-0.45,-0.5);
\draw (-3.7,-2.1)--(-0.45,-3);

\node at (-3,-5.5) 
{
\hspace{-1cm}
\includegraphics[scale=0.33]{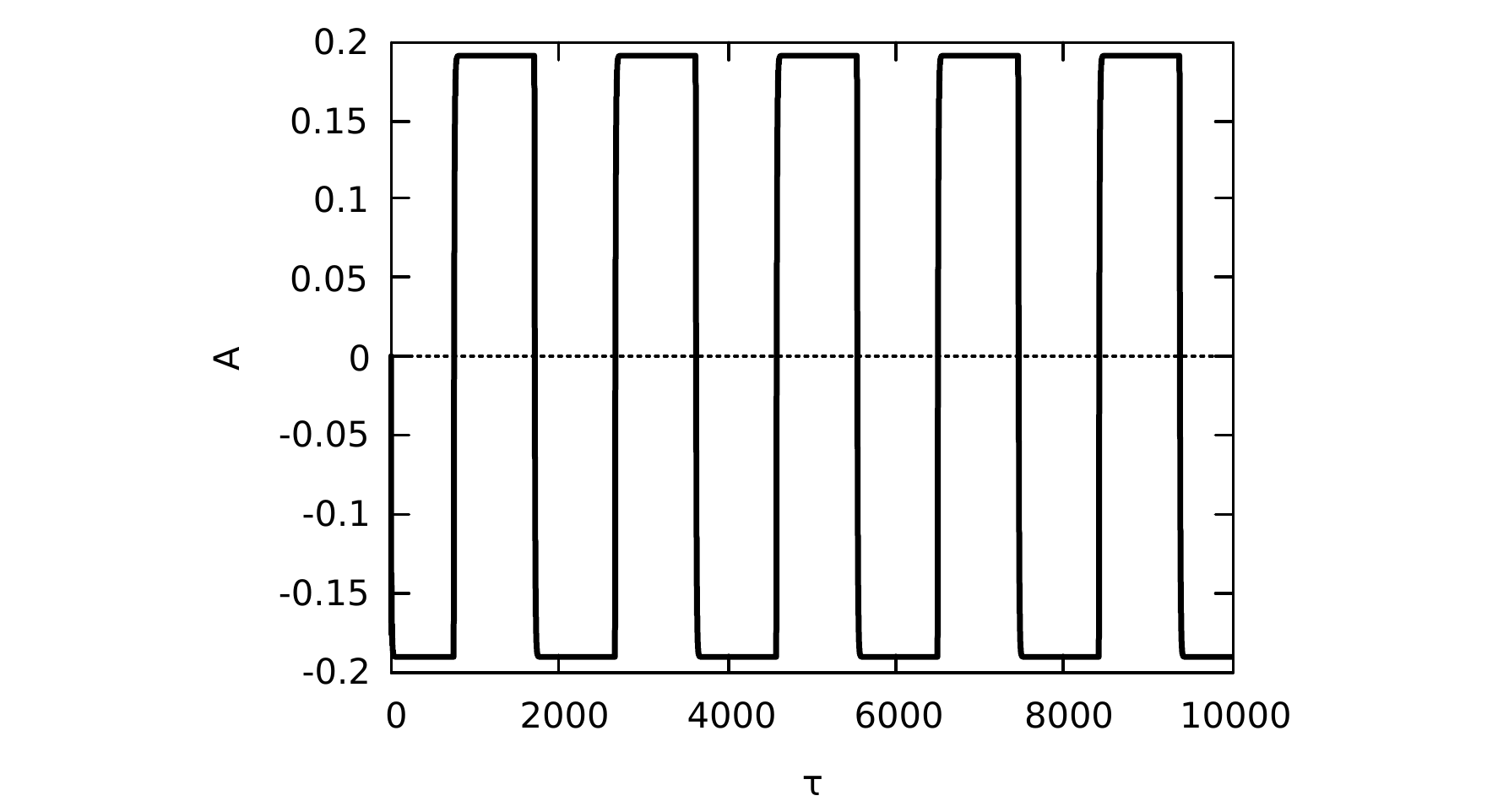}
\hspace{-1.75cm}
\includegraphics[scale=0.33]{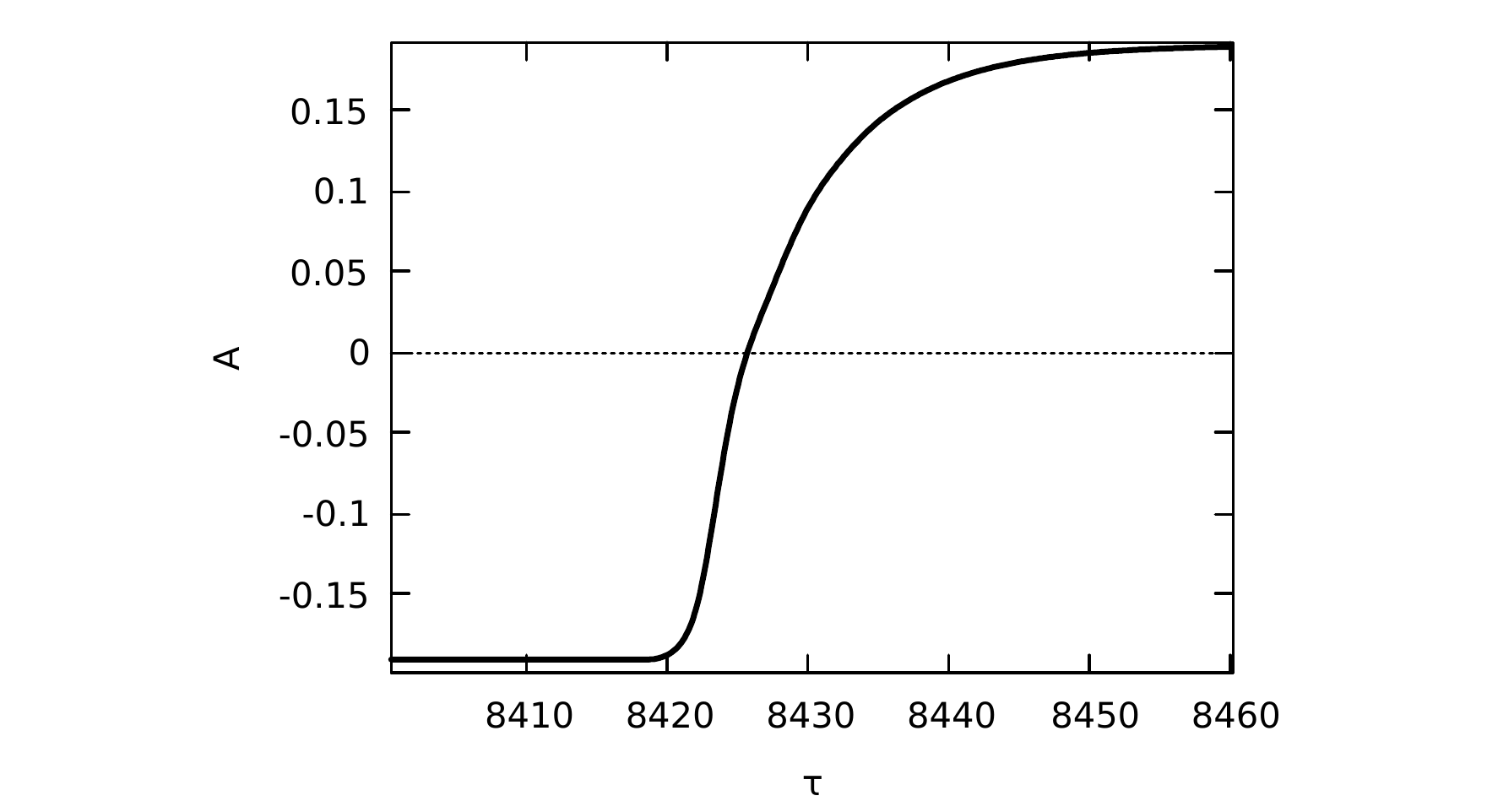}
\hspace{-1.75cm}
\includegraphics[scale=0.33]{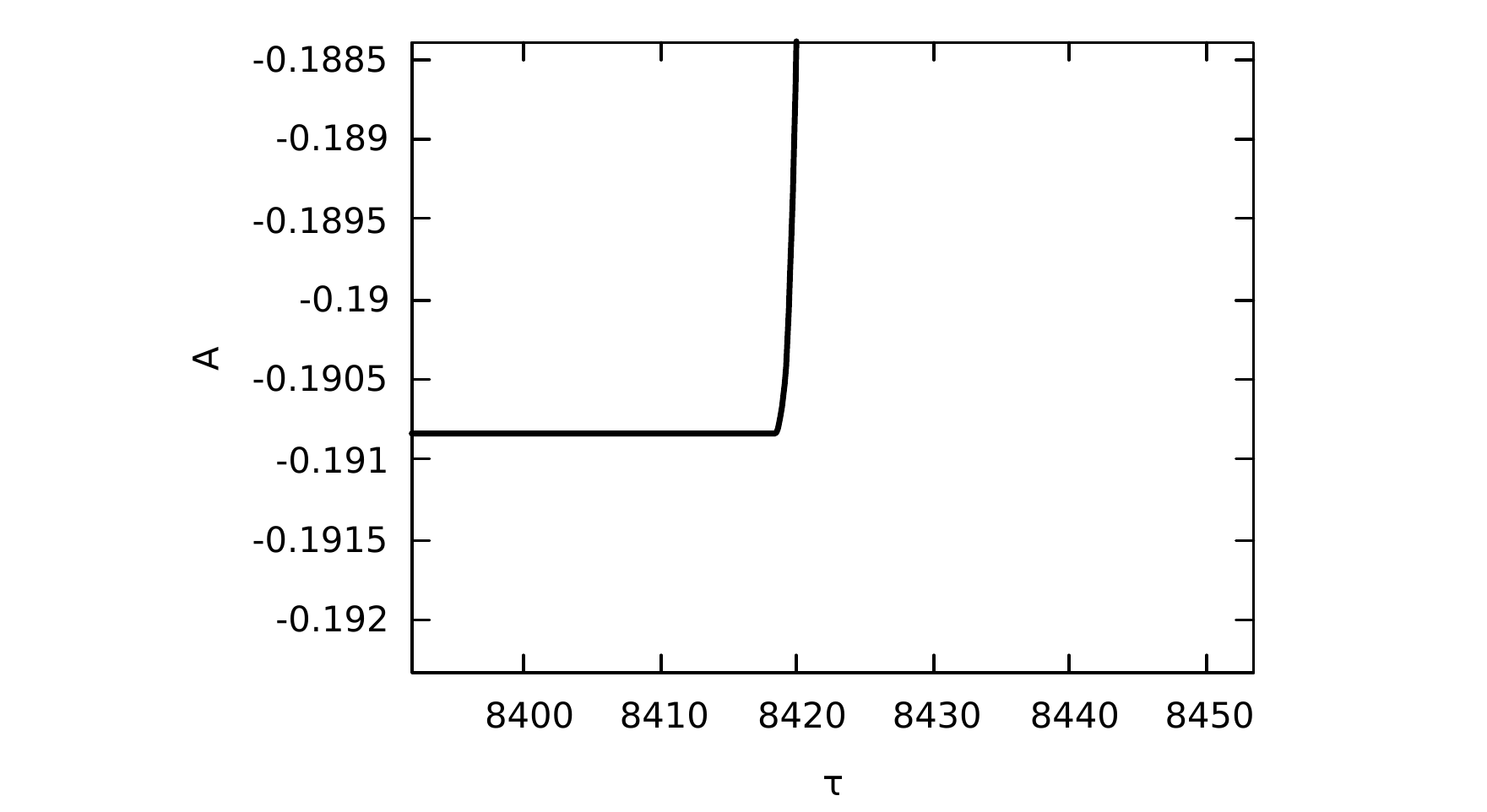}
};

\draw(-6.45,-5.3) ellipse (0.1cm and 1.6cm);
\draw (-6.45,-3.7)--(-4.8,-4.1);
\draw (-6.45,-6.9)--(-4.8,-6.6);

\draw(-3.7,-6.5) ellipse (0.2cm and 0.3cm);
\draw (-3.7,-6.2)--(-0.41,-4.1);
\draw (-3.7,-6.8)--(-0.41,-6.6);

\node at (-3,-9) 
{
\hspace{-1cm}\includegraphics[scale=0.33]{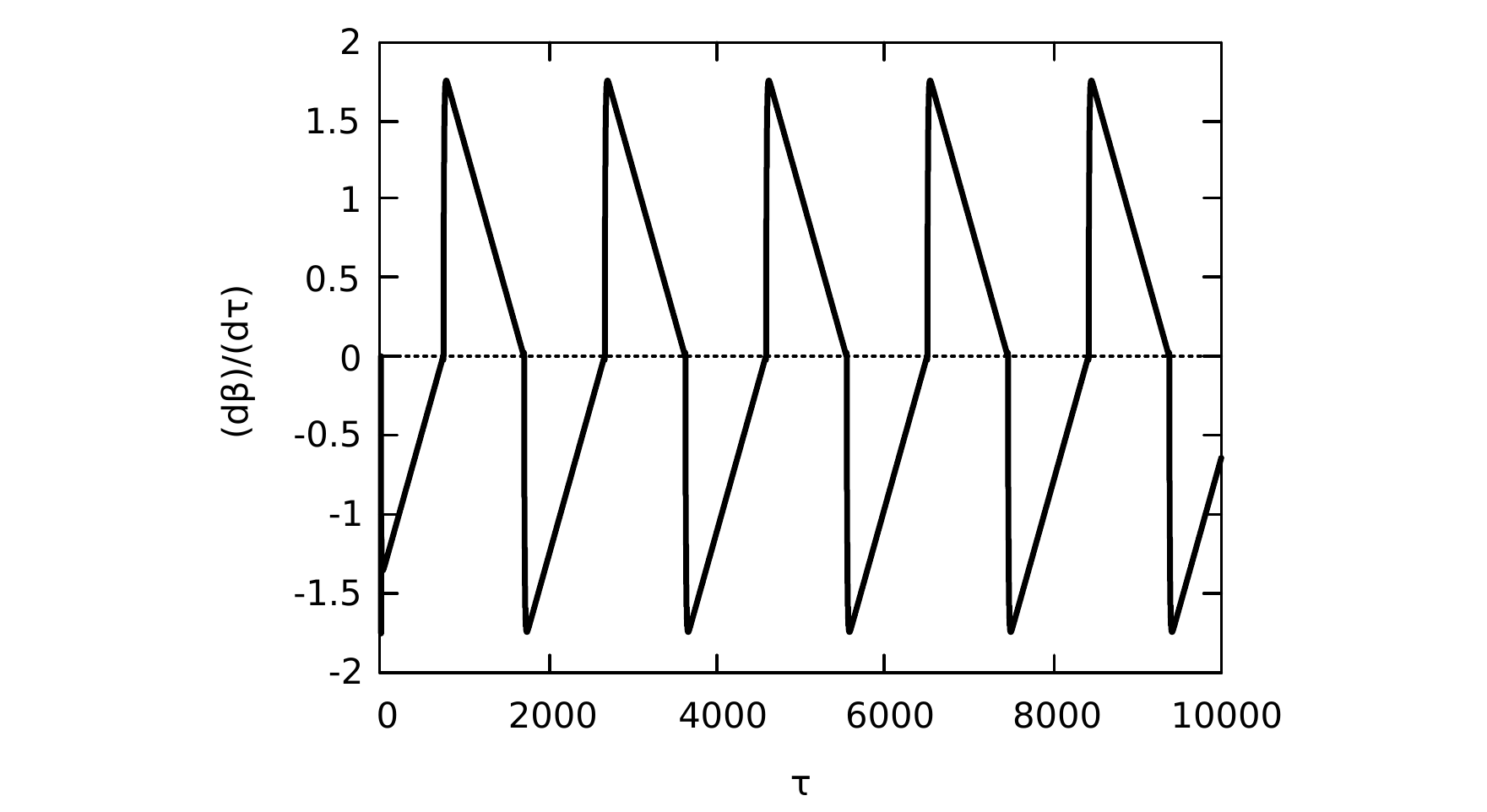}
\hspace{-1.75cm}\includegraphics[scale=0.33]{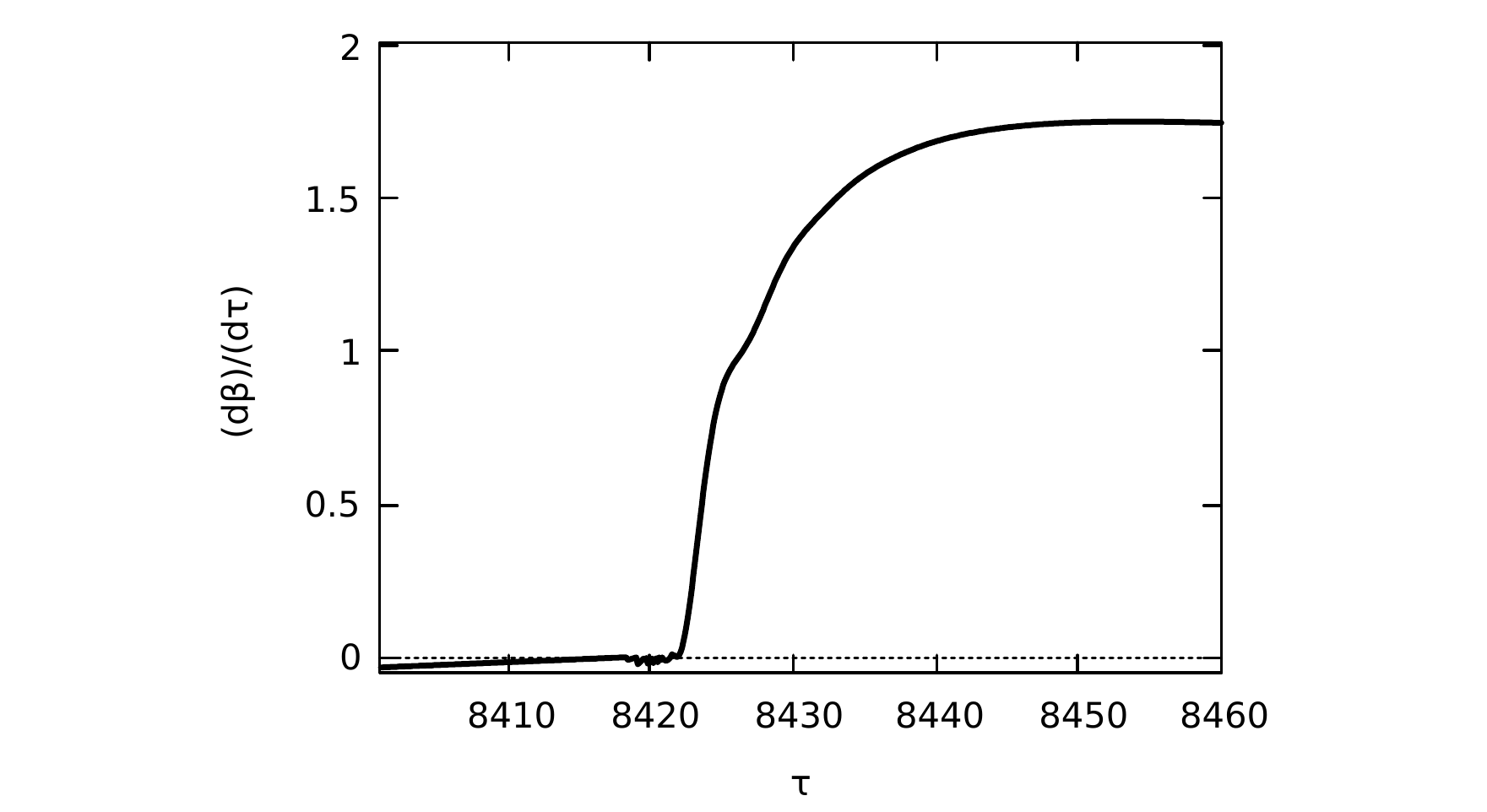}
\hspace{-1.75cm}
\includegraphics[scale=0.33]{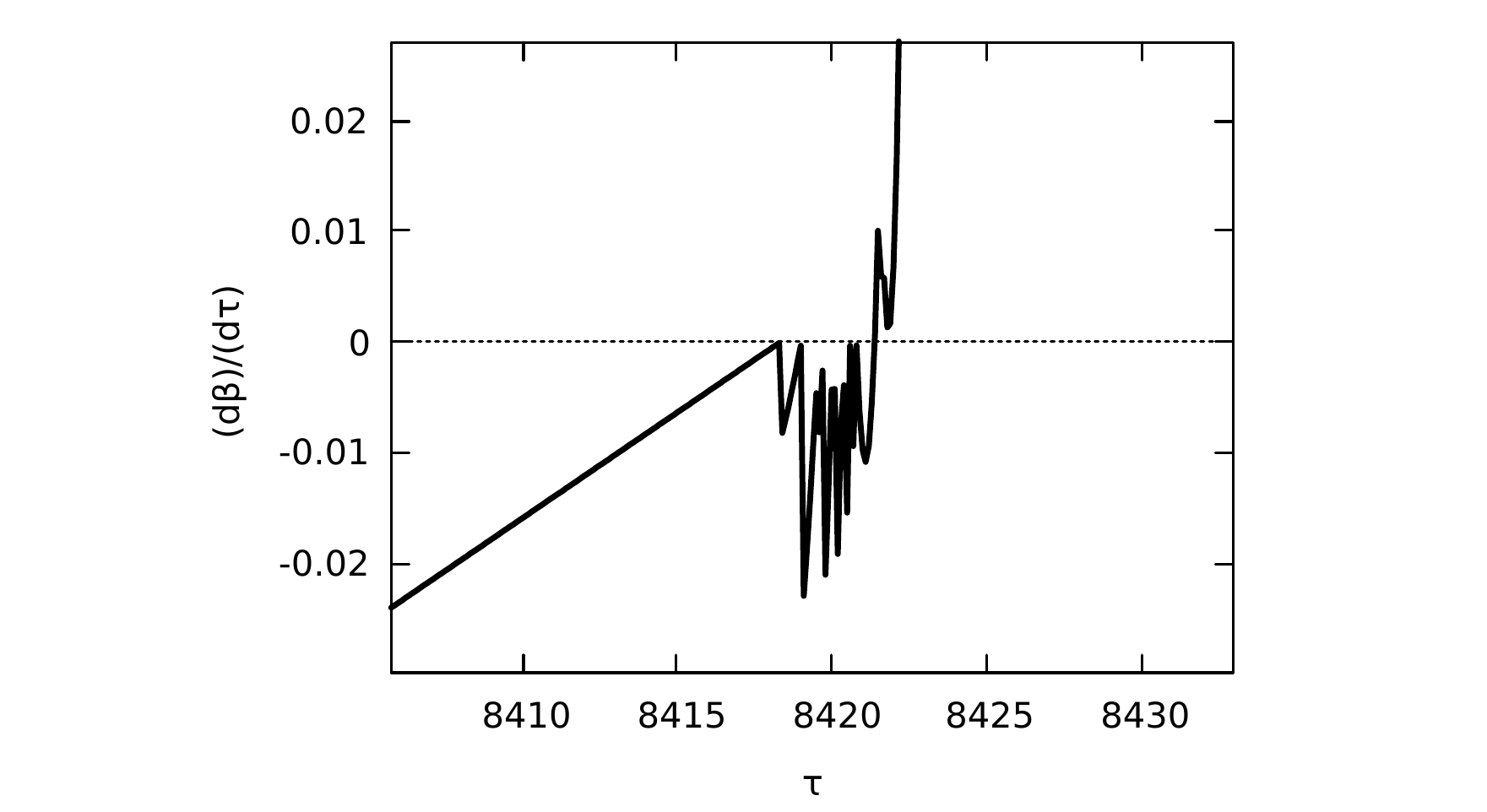}};

\draw(-6.5,-8.4) ellipse (0.1cm and 0.8cm);
\draw (-6.45,-7.6)--(-5,-7.5);
\draw (-6.45,-9.2)--(-5,-10);

\draw(-3.9,-9.8) ellipse (0.2cm and 0.3cm);
\draw (-3.9,-9.5)--(-0.47,-7.5);
\draw (-3.9,-10)--(-0.47,-10);

\end{tikzpicture}

\caption{
Here one can see the solution of (\ref{eqForPendulumOnWheelFullWithStabAndDump}) with the  initial conditions $\tau=0$, $\alpha=-0.1,\dot{\alpha}=-0.1,\beta=0,\dot{\beta}=0$ and under the proportional-integral-derivative controller, where $k_1=1.5,k_2=0.2,k_3=0.05$. On the upper figure one can see stabilization of $\alpha$; on the second row of the figures it is shown   $\dot{\alpha}$; on the third row of the figures one can see  $A=\int_0^\tau \alpha(\theta)d\theta$; on the lowest row of the figures it is shown  $\dot{\beta}(\tau)$. The right column of the figures show the behaviour of the numerical solution near the unstable point. The parameters of the WIP are following: $\zeta=10,\,\rho=0.2$,  $\nu=0.1$.}
\label{figPIDController}
\end{figure}

\begin{figure}
\includegraphics[scale=0.4]{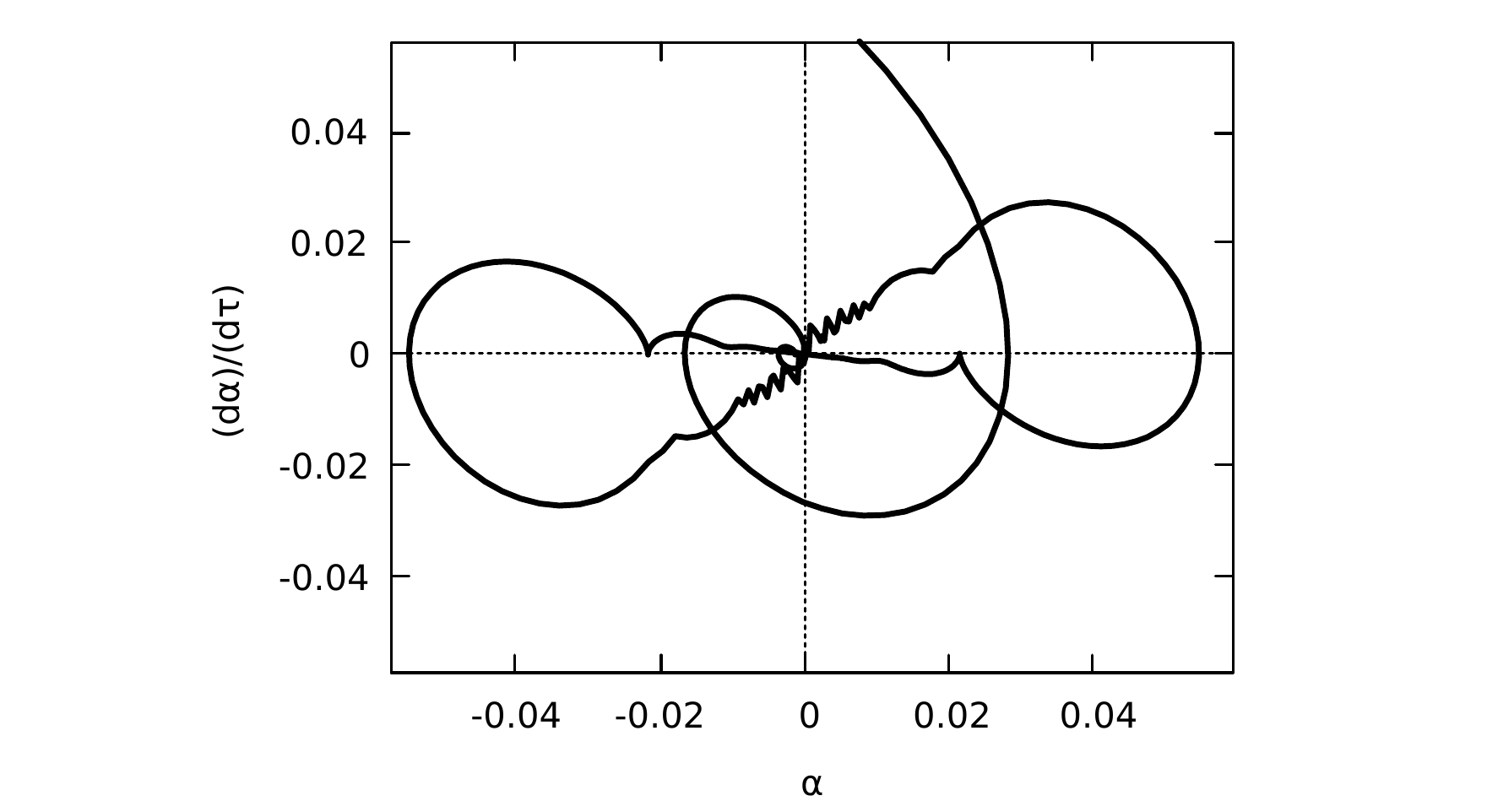}
\includegraphics[scale=0.4]{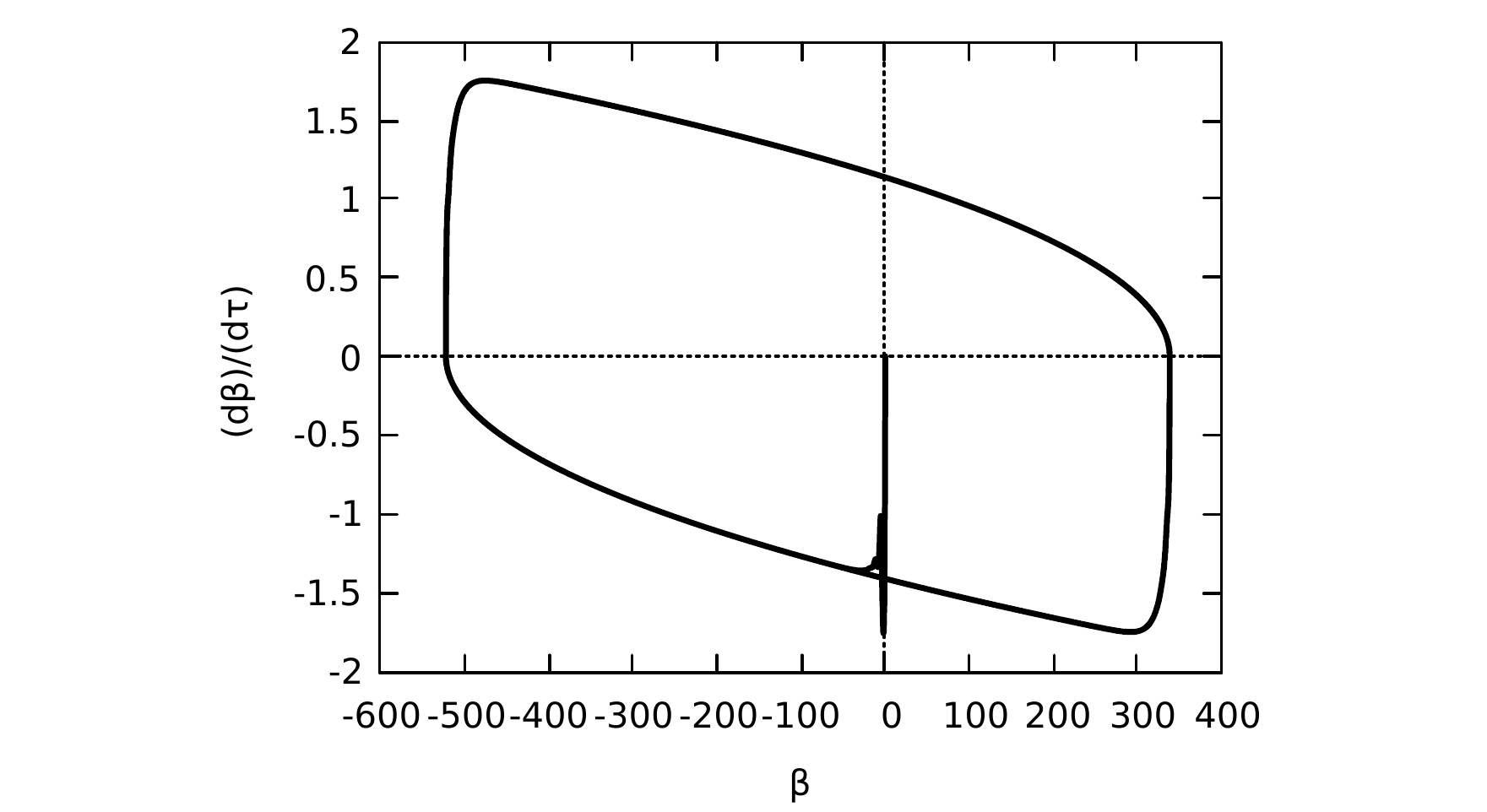}
\caption{
Here one can see the solution of  (\ref{eqForPendulumOnWheelFullWithStabAndDump}) with the  initial conditions $\tau=0$, $\alpha=-0.1,\dot{\alpha}=-0.1,\beta=0,\dot{\beta}=-0.1$ and under the proportional-integral-derivation controller, where $k_1=1.5,k_2=0.2,k_3=0.05$. On the left-hand side  it is shown the limit cycle on the phase plane $(\alpha,\dot{\alpha})$ and on the right-hand side it is shown the limit cycle on the phase plane $(\beta,\dot{\beta})$.  The parameters of the pendulum are following: $\zeta=10,\,\rho=0.2$ and $\nu=0.1$.}
\label{figLimitCycleForWheel}
\end{figure}

Let us consider the proportional-integral-derivative controller to stabilize the WIP:
$$
h=k_1\alpha+k_2\dot{\alpha}+k_3\int_0^\tau\alpha(\theta)d\theta.
$$
The integral term in the control adds an additional term which is linearly growing for all  $\alpha=\const\not=0$. Therefore there exists only one stationary solution $\alpha\equiv0$.  Really, let us denote
$$
A=\int_0^\tau \alpha d\theta,
$$
then we obtain the third order equation:

\begin{eqnarray*}
(2+\zeta\sin^2(\dot{A}))\dddot{A}=
&
(2+\zeta)\sin(\dot{A})-\frac{\zeta}{2}\ddot{A}^2\sin(2\dot{A})
+\sgn(\dot{\beta})\nu\cos(\dot{A})-
\\
&
2\rho\left(1+\frac{2}{\zeta}\right) (k_1\dot{A}+k_2\ddot{A}+k_3A)-
\\
&
\frac{2}{\rho}\cos(\dot{A})(k_1\dot{A}+k_2\ddot{A}+k_3 A).
\end{eqnarray*}

The stationary solution of this equation as  $\alpha=\dot{A}\equiv0$) is following:
$$
A^\pm=\frac{\operatorname{sgn}(\dot{\beta})  \zeta  \nu  \rho }
{(2k_3 \zeta +4 k_3) \, \rho^2+2 k_3\zeta }.
$$

Define:
$$
h=k_1\alpha+k_2\dot{\alpha}+A^\pm+k_3\int_{\tau^\pm}^\tau \alpha d\theta.
$$
When $\dot{\beta}\not=0$ and  $\tau>\tau^\pm$  we get the equation for dynamics on the horizontal with the integral term:
\begin{eqnarray}
(\sin^2(\alpha)\zeta +2)\ddot{\alpha}
=(\zeta +2)\sin( \alpha)- \frac{1}{2}\dot{\alpha}^2\zeta\sin(2\alpha)-
\nonumber\\
2\left(\frac{1}{\rho}\cos(\alpha)+\left(1+\frac{2}{\zeta}\right)\rho \right)(k_1\alpha+k_2\dot{\alpha}+k_3\int_{\tau^\pm}^\tau \alpha d\theta.).
\label{EqPendulumOnWheel1WithIntegral}
\end{eqnarray}
If $k_2=k_3=0$, then the conservation law coincides with the conservation law for the WIP on the horizontal:
\begin{eqnarray*}
\mathcal{E}_0
&=
\left(1+\frac{1}{2}\sin^2(\alpha)\zeta\right)\dot{\alpha}^2+(\zeta+2)\cos(\alpha)+
\\
&
k_1\left(\frac{2}{\rho}(\alpha\sin(\alpha)+ \cos(\alpha)+\alpha^2\rho\left(1+\frac{2}{\zeta}\right)\right).
\end{eqnarray*}
Let us differentiate $\mathcal{E}_0$ when  $\sgn(\dot{\beta})=\const$:  
$$
\frac{d\mathcal{E}_0}{d\tau}=-2k_2\left(\left(1+\frac{2}{\zeta}\right)\rho+\frac{1}{\rho}\cos(\alpha)\right)\left(\dot{\alpha}^2+\kappa_3\dot{\alpha}\int_{\tau^\pm}^\tau \alpha(\theta) d\theta\right).
$$
Here $\kappa_3=k_3/k_2$.

If $\kappa_3=0$, then the movement coincides with the movement on the horizontal which is stable (see theorem \ref{theoremAboutStabPDOnHorizontal}). If $\kappa_3>0$, then the right-hand side of the formula contains the integral term in which sign is not defined. But for small values of $\kappa$ one can hope, that the integral term do not leads to the instability of $\alpha\equiv0$.

To study the stability for small $\alpha$ we consider the linearized equation (\ref{EqPendulumOnWheel1WithIntegral}) near $(\alpha,\dot{\alpha})=(0,0)$:
$$
\ddot{a}=-k_3s\int_{\tau^\pm}^\tau a(\theta)d\theta
-k_2s\dot{a}-\left(s-\frac{\zeta }{2}-1\right)a,
$$
where
$$
s=\left(\frac{2\rho}{\zeta }+\rho +\frac{1}{\rho }\right).
$$
Let us differentiate this equation with respect to $\tau$. Then the characteristic equation has the form:
$$
\lambda^3=-k_3s
-k_2s\lambda^2-\left(s-\frac{\zeta }{2}-1\right)\lambda.
$$
Let us denote the solutions of this equation as
$\lambda_j,\, j=1,2,3$.
In general case the condition of the stability for $(\alpha,\dot{\alpha})=(0,0)$ is following:
$$
\Re(\lambda_j)<0,\quad j=1,2,3.
$$
If $k_2$ and $k_3$ are small, then the condition of the stability has the form:
\begin{equation}
\left(\frac{2\rho}{\zeta }+\rho+\frac{1}{\rho}\right)k_1 > \frac{\zeta }{2}+1,\quad
0<k_3<k_2 \left(\frac{2\rho}{\zeta }+\rho +\frac{1}{\rho }\right).
\label{conditionsForStabilityPID}
\end{equation}

Let us formulate the main result.
\begin{theorem}\label{theoremAboutStabPIDOnSoftSurface}
There exist $k_2^0>0$ and $k_3^0>0$ such, that the equation  (\ref{EqPendulumOnWheel1WithIntegral}) as $\sgn(\dot{\beta})=\const$ have the asymptotic stable equilibrium point $(\alpha_0,0)$ on the phase plane as (\ref{conditionsForStabilityPID}), where $k_2<k_2^0$ and $k_3<k_3^0$.
\end{theorem}

The theorem \ref{theoremAboutStabPIDOnSoftSurface} allows us to obtain the quality properties for the dynamics of the WIP  under the proportional-integral-derivative controller  on the soft surface. 

The typical phases of the movement one can see on the figures \ref{figPIDController}. These figures show the numerical solutions of the differential inclusions for the WIP on the soft surface. In particular   the third row of the figures show the fast growth of the integral term in the controller, then this term stabilises while the $\dot{\beta}$ is not change their sign. In the region with the stable value of the integral term we see  $(\alpha,\dot{\alpha})\to(0,0)$ and $\dot{\beta}\to-0$. 

The limit value of  $A=A^{-},\alpha=0,\dot{\alpha}=0, \dot{\beta}=0$ is the stationary solution which is stable for $\dot{\beta}\to-0$. But this stationary solution is not stable for  $\dot{\beta}=0+\delta$ $\forall\delta>0$. For real robotics equipments there are the unused perturbations in the mathematical model or the rounding errors in the numerical solutions. These perturbations and errors lead to change the sign of $\dot{\beta}$ due to the instability.

The system fast changes after the change of the sign of $\dot{\beta}$. The integral term stabilises at the new value $A=A^{+}$. The the process of stabilization repeats and  $(\alpha,\dot{\alpha})\to(0,0)$ and $\dot{\beta}\to+0$. In this case the limit value $A=A^{+},\alpha=0,\dot{\alpha}=0, \dot{\beta}=0$  and we get the stationary solution which is stable for $\dot{\beta}\to+0$ but not stable for $\dot{\beta}\to-0$. 

Over time due to the perturbations in the robotic equipment or rounding errors in the numeric solution  the $\dot{\beta}$ changes the sign and the process repeated. 

This is the scheme of an appearance for the limit cycle. The period of this cycle  depends on a random value.

{\bf Important note.} The limit cycle appears for the real robotic equipment or for the  numerical solutions due to the semi-stability of the solutions $\alpha\equiv0,\dot{\beta}\equiv0,A\equiv A^{\pm}$. Pure analytical solutions without any additional perturbations and any rounded errors should stabilize to $\alpha\equiv0,\dot{\beta}\equiv0,A\equiv A^{\pm}$. The right column of the figures shows the behaviour of the numerical solution near the semi-stable solution.

On the figure \ref{figLimitCycleForWheel} it is shown the projections of the trajectories onto two phase planes  $(\alpha,\dot{\alpha})$ and $(\beta,\dot{\beta})$.

\section{Conclusions}

It is shown that to stabilize the pendulum on the wheel on a flat and inclined surface, it is sufficient to use the proportional-derivative controller.

To stabilize the pendulum on the wheel on a soft surface we show, that   the proportional-derivative controller can stabilize the pendulum in the upper position, however, the wheel moves with a constant acceleration. Such acceleration is not acceptable for the robotic equipments. Therefore to stabilize the wheel rotation, it is proposed to use the proportional-integral-derivative controller.

It turns out that the stabilization on the soft surface leads to the limit cycle near the state of the unstable equilibrium of the pendulum for the numeric solutions and the physical robotic equipment. The period of this limit cycle depends on a random variable.

\end{document}